%% file: LP_v27_arXiv_v3.tex
\pdfoutput=1
\PassOptionsToPackage{
pdfencoding=auto,
pdfnewwindow=true,
pdfusetitle=false,
bookmarks=true,
bookmarksnumbered=true,
bookmarksopen=true,
pdfpagemode=UseThumbs,
bookmarksopenlevel=1,
pdfpagelabels=false,
breaklinks=true,  
backref=false,
colorlinks=true,
}{hyperref}
\RequirePackage{etex}
\RequirePackage{fix-cm}
\PassOptionsToPackage{skip=0pt, font=small,labelfont=bf}{caption}
\documentclass[10pt,showpacs,letterpaper,aps,pra,twocolumn,nofootinbib,superscriptaddress,floatfix]{revtex4-2} 
\usepackage{lmodern}
\usepackage{xfrac}

\usepackage{paralist}

\usepackage[toc,page,header]{appendix}
\usepackage{minitoc}

\makeatletter
\renewcommand*\l@section{\@dottedtocline{1}{0.0em}{1.0em}}
\renewcommand*\l@subsection{\@dottedtocline{2}{1.0em}{1.0em}}
\renewcommand*\l@subsubsection{\@dottedtocline{3}{2.0em}{1.0em}}
\makeatother

\usepackage{hyperref}
\usepackage{amsthm,amsmath,amssymb,graphicx,comment,dsfont} 
\usepackage{color}
\usepackage{bbold}
\DeclareMathAlphabet{\mathfrak}{OMS}{ygoth}{m}{n}

\usepackage[none]{hyphenat}
\usepackage{fancyhdr}
\newtheorem{theorem}{Theorem}

\newtheorem{form}{Formulation}

\newtheorem{lp}{Linear Program}

\newtheorem{prop}{Proposition}

\newtheorem{lemma}[theorem]{Lemma}

\newtheorem{definition}[theorem]{Definition}

\newcommand{\ket}[1]{\left| #1 \right\rangle}
\newcommand{\bra}[1]{\left\langle #1 \right|}

\newcommand{\beq}{\begin{equation}}
\newcommand{\eeq}{\end{equation}}
\newcommand{\bea}{\begin{align}}
\newcommand{\eea}{\end{align}}

\newcommand{\discard}{\resizebox{4mm}{!}{%
\InputIfFileExists{Diagrams/discardGPT.tikz}{}{\input{./figures/Diagrams/discardGPT.tikz}}}}
\newcommand{\discardA}{\resizebox{4mm}{!}{%
\InputIfFileExists{Diagrams/discard.tikz}{}{\input{./figures/Diagrams/discard.tikz}}}}

\newcommand{\gse}{\tau_{\Omega^\mathcal{G}}}
\newcommand{\gee}{\tau_{\mathcal{E}^\mathcal{G}}}
\newcommand{\fse}{\tau_{\Omega^\mathfrak{F}}}
\newcommand{\fee}{\tau_{\mathcal{E}^\mathfrak{F}}}
\newcommand{\ase}{\tau_{\Omega^\mathfrak{A}}}
\newcommand{\aee}{\tau_{\mathcal{E}^\mathfrak{A}}}

\usepackage{graphicx}
\usepackage[usenames,dvipsnames,table]{xcolor} 
\usepackage{mathtools}
\usepackage[normalem]{ulem}
\input{preamble}

\definecolor{googleblue}{RGB}{34, 0, 204}
\definecolor{panblue}{RGB}{0,24,150}
\definecolor{carmine}{RGB}{150, 0, 24}
\hypersetup{
unicode=true,
bookmarksnumbered=false,
bookmarksopen=false,
breaklinks=false,
colorlinks=true,
linkcolor=carmine,
citecolor=googleblue,
urlcolor=panblue,
anchorcolor=OliveGreen}

\DeclareRobustCommand*{\blk}{}
\newcommand{\john}{}

\allowdisplaybreaks

\usepackage{subcaption}
\usepackage[labelformat=simple]{subcaption}

\usepackage{microtype}
\microtypecontext{spacing=nonfrench}
\microtypesetup{
expansion={true,nocompatibility},
protrusion={true,nocompatibility},
activate={true,nocompatibility},
tracking=true,
kerning=true,
spacing={true}
}

\usepackage[all=normal,floats=tight,mathspacing=tight,wordspacing=tight,paragraphs=normal,tracking=tight,charwidths=tight,mathdisplays=normal,sections=normal,margins=normal]{savetrees}

\everypar=\expandafter{\the\everypar\loosness=-1 }
\newcommand{\nocontentsline}[3]{}
\let\oldaddcontentsline\addcontentsline
\newcommand{\tocless}[2]{%
  \let\addcontentsline=\nocontentsline#1{#2}
  \let\addcontentsline\oldaddcontentsline}

\newcommand{\n}{n}
\newcommand{\m}{m}
\usepackage{bbm}

\usepackage{tikz}
\usetikzlibrary{positioning,backgrounds}
\usetikzlibrary{calc,intersections,through}
\definecolor{darkgreen}{rgb}{0,.5,0}
\definecolor{darkblue}{rgb}{0,0,.5}
\definecolor{darkred}{rgb}{0.5,0,0}

\begin{document}

\doparttoc 
\faketableofcontents 

\part{} 

\title{A linear program for testing nonclassicality and an open-source implementation}

\author{John H.~Selby}
\email{john.h.selby@gmail.com}
\affiliation{International Centre for Theory of Quantum Technologies, University of Gda\'nsk, 80-309 Gda\'nsk, Poland}
\author{Elie Wolfe}
\email{ewolfe@perimeterinstitute.ca}
\affiliation{Perimeter Institute for Theoretical Physics, 31 Caroline Street North, Waterloo, Ontario Canada N2L 2Y5}
\author{David Schmid}
\email{davidschmid10@gmail.com}
\affiliation{International Centre for Theory of Quantum Technologies, University of Gda\'nsk, 80-309 Gda\'nsk, Poland}
\author{Ana Bel\'en Sainz}
\email{ana.sainz@ug.edu.pl}
\affiliation{International Centre for Theory of Quantum Technologies, University of Gda\'nsk, 80-309 Gda\'nsk, Poland}
\author{Vinicius P. Rossi}
\email{prettirossi.vinicius@gmail.com}
\affiliation{International Centre for Theory of Quantum Technologies, University of Gda\'nsk, 80-309 Gda\'nsk, Poland}

\date{\today}

\begin{abstract}
		A well motivated method for demonstrating that an experiment resists any classical explanation is to show that its statistics violate generalized noncontextuality.
		We here formulate this problem as a linear program and provide an open-source implementation of it which tests whether or not any given prepare-measure experiment is classically-explainable in this sense. The input to the program is simply an arbitrary set of quantum states and an arbitrary set of quantum effects; the program then determines if the Born rule statistics generated by all pairs of these can be explained by a classical (noncontextual) model. If a classical model exists, it provides an explicit model. If it does not, then it computes the minimal amount of noise that must be added such that a model does exist, and then provides this model. 
		We generalize all these results to arbitrary generalized probabilistic theories (and accessible fragments thereof) as well; indeed, our linear program is a test of simplex-embeddability as introduced in Ref.~\cite{schmid2021characterization} and generalized in Ref.~\cite{selby2021accessible}. 
\end{abstract}
\maketitle

	A rigorous method for demonstrating that a theory or a set of data resists any classical explanation is to prove that it cannot be reproduced in any generalized noncontextual model~\cite{gencontext}. Generalized noncontextuality was first introduced as an improvement on Kochen-Specker's assumption of noncontextuality~\cite{KS}, making it more operationally accessible and providing stronger motivations for it, as a form of Leibniz's principle~\cite{Leibniz}. Since its inception, the list of motivations for taking it as one's notion of classicality has grown greatly.
	Notably, the existence of a generalized-noncontextual ontological model for an operational theory coincides with two independent notions of classicality: one that arises in the study of generalized probabilistic theories~\cite{schmid2021characterization,shahandeh2021contextuality,schmid2020structure}, and another that arises in quantum optics~\cite{negativity,schmid2021characterization,schmid2020structure}. 
	Generalized noncontextuality has been used as an indicator of classicality in the quantum Darwinist program~\cite{baldijao2021noncontextuality}, and any sufficiently noisy theory satisfies generalized noncontextuality~\cite{operationalks,marvian2020inaccessible}. Furthermore, violations of local causality~\cite{Bell}, violations of Kochen-Specker noncontextuality~\cite{operationalks,kunjwal2018from}, and some observations of anomalous weak values~\cite{AWV, KLP19}, are also instances of generalized contextuality. Finally, generalized contextuality is a resource for information processing~\cite{POM,RAC,RAC2,Saha_2019,YK20}, computation~\cite{schmid2021only}, state discrimination~\cite{schmid2018contextual,flatt2021contextual,mukherjee2021discriminating,Shin2021}, cloning~\cite{cloningcontext}, and metrology~\cite{contextmetrology}.  Herein,  we use  the term \textit{noncontextuality} to refer to the concept of generalized noncontextuality.
	
	How, then, does one determine in practice whether a given theory or a given set of experimental data admits of a classical explanation of this sort? We here provide the most direct algorithm to date for answering this question in arbitrary prepare-and-measure experiments, and we provide open-access Mathematica code for answering it in practice. {\bf One need only give a  finite  set of quantum states and a  finite  set of quantum POVM elements as input, and the code determines if the statistics these generate by the Born rule can be explained classically---i.e., by a noncontextual ontological model for the operational scenario.} It furthermore returns an explicit noncontextual model, if one exists. If there is no such model, the code determines an operational measure of nonclassicality, namely, the minimum amount of noise which would be required until a noncontextual model would become possible.
	
	In the Supplemental Material~\cite{footnote1}, we generalize these ideas beyond quantum theory to the case of arbitrary  generalized probabilistic theories (GPTs) \cite{hardy2001quantum,GPT_Barrett} or fragments thereof, leveraging the fact that an operational scenario admits of a noncontextual model if and only if the corresponding GPT admits of a simplex-embedding~\cite{schmid2021characterization}. Indeed, the linear program we derive is simply a test of whether any valid simplex-embedding (of any dimension) can be found, answering the challenge first posed in Ref.~\cite{schmid2021characterization}. We furthermore prove an upper bound on the number of ontic states needed in any such a classical explanation, namely, the square of the GPT dimension. 
	
	The Supplemental Material also explains how our open-source code implements the linear program we develop herein.
	
	A large number of previous works have studied the question of when a set of data admits of a generalized noncontextual model~\cite{merminsquare,Schmid2018,robust,schmid2021characterization,gitton2020solvable,shahandeh2021contextuality,schmid2020structure,selby2021accessible,selby2021incompatibility,mueller2021testing,Chaturvedi2021characterising}. Most closely related to our work are Refs.~\cite{merminsquare,Schmid2018,gitton2020solvable,shahandeh2021contextuality}. We elaborate on the relationships between these works in our conclusion and in our Supplemental Material. 
	
	For now, we simply note that the linear program (and dimension bound) that we derive here is closely related to an optimization problem introduced in Ref.~\cite{gitton2020solvable}. 
	However, Ref.~\cite{gitton2020solvable} focuses on a proposed modification of generalized noncontextuality  (which we criticize in the Supplemental Material),  and so the two approaches do not always return the same result.

	Our manuscript aims to be accessible and self-contained, in order to provide a tool for the quantum information and foundations communities to directly test for nonclassicality in their own research problems. 
	
	\noindent\emph{A linear program for deciding classicality---}
	We now set up the preliminaries required to state our linear program for testing whether the quantum statistics generated by given sets  of quantum states and effects can be explained classically---i.e., by a noncontextual model  for the operational scenario. The Supplemental Material generalizes these ideas and results to arbitrary GPTs.
	
	Consider any  finite set of (possibly subnormalized\footnote{This allows one to describe states which are not prepared deterministically, as happens, e.g., when the preparation is part of a probabilistic source or is the result of remote steering.}) quantum states, $\Omega$, and any  finite  set of quantum effects, $\mathcal{E}$, living in the real vector space $\mathsf{Herm}[\mathcal{H}]$ of Hermitian operators on some  finite dimensional  Hilbert space $\mathcal{H}$. In general, neither the set of states nor the set of effects need span the full vector space $\mathsf{Herm}[\mathcal{H}]$, nor need the two sets span the same subspace of $\mathsf{Herm}[\mathcal{H}]$. 
	Next, we introduce some useful mathematical objects related to $\Omega$ and $\mathcal{E}$.

	Let us first focus on the case of states. 
	We denote the subspace of $\mathsf{Herm}[\mathcal{H}]$ spanned by the states $\Omega$ by $S_\Omega$.
	The inclusion map from $S_\Omega$ to $\mathsf{Herm}[\mathcal{H}]$ is denoted by $I_\Omega$. 
	In addition, we define the cone of positive operators that arises from $\Omega$  by 
	\beq
	\mathsf{Cone}[\Omega]=\left\{\rho \middle| \rho = \sum_{\alpha} r_\alpha \rho_\alpha, \rho_\alpha \in \Omega, r_\alpha \in \mathds{R}^+ \right\}  \subset S_\Omega\,.
	\eeq
	\noindent This cone can also be characterized by its facet inequalities, indexed by $i=\{1,...,n\}$,  where $n$ is necessarily finite as we start with a finite set of states (see, for example, McMullen's upper bound theorem \cite{mcmullen1970maximum}).   These inequalities are  specified by Hermitian operators $h_i^\Omega \in S_\Omega$ such that
	\beq \label{eq:statefacets}
	\mathsf{tr}(h_i^\Omega v) \geq 0\ \ \forall i \quad \iff \quad v \in \mathsf{Cone}[\Omega].
	\eeq
	From these facet inequalities, one can define a linear map $H_\Omega:S_\Omega \to \mathds{R}^n$, such that
	\beq
	H_\Omega(v) = \left(\mathsf{tr}(h_1^\Omega v) ,...,\mathsf{tr}(h_n^\Omega v)\right)^T \quad \ \forall v\in S_\Omega.
	\eeq 
	Note that the matrix elements of $H_\Omega(v)$ are all non-negative if and only if $v\in \mathsf{Cone}[\Omega]$. We denote entrywise non-negativity by $H_\Omega(v)\geq_e 0$ (to disambiguate from using $\geq 0$ to represent positive semi-definiteness). Succinctly, we have
	\beq
	H_\Omega(v) \geq_e 0 \ \ \iff \ \ v\in\mathsf{Cone}[\Omega],
	\eeq
	and so $H_\Omega$ is simply an equivalent characterisation of the cone.
	
	Consider now the set of effects $\mathcal{E}$. We denote the subspace of $\mathsf{Herm}[\mathcal{H}]$ spanned by $\mathcal{E}$ by $S_\mathcal{E}$, and the inclusion map from $S_\mathcal{E}$ to $\mathsf{Herm}[\mathcal{H}]$ by $I_\mathcal{E}$.
	In addition, we define the cone of positive operators that arises from $\mathcal{E}$  as 
	\beq
	\mathsf{Cone}[\mathcal{E}]=\left\{\gamma\middle|\gamma=\sum_\beta r_\beta \gamma_\beta, \gamma_\beta\in \mathcal{E}, r_\beta\in\mathds{R}^+\right\} \subset S_\mathcal{E}
	\eeq
	\noindent This cone can also be characterised by its facet inequalities, indexed by $j=\{1,...,m\}$,  where $m$ is again finite, as we are considering a finite set of effects.  These inequalities are  specified by Hermitian operators $h_j^\mathcal{E}$ such that
	\beq \label{eq:effectfacets}
	\mathsf{tr}(h_j^\mathcal{E} w) \geq 0 \ \forall j\quad \iff\quad w \in \mathsf{Cone}[\mathcal{E}].
	\eeq
	From these facet inequalities one can define a linear map $H_\mathcal{E}:S_\mathcal{E}\to \mathds{R}^m$, such that
	\beq
	H_\mathcal{E}(w)= \left(\mathsf{tr}(w h_1^\mathcal{E}),...,\mathsf{tr}(wh_m^\mathcal{E})\right)^T\ \quad \forall w\in S_\mathcal{E}.
	\eeq
	This fully characterises $\mathsf{Cone}[\mathcal{E}]$, since
	\beq
	H_\mathcal{E}(w)\geq_e 0 \quad \iff\quad w\in\mathsf{Cone}[\mathcal{E}].
	\eeq

	One can also pick an arbitrary othonormal basis of Hermitian operators for each of the spaces $\mathsf{Herm}[\mathcal{H}]$,  $S_\Omega$, and  $S_\mathcal{E}$, and represent $I_\Omega$, $I_\mathcal{E}$,  $H_\mathcal{E}$, and $H_\Omega$ as matrices with respect to these.
	
	With these defined, we can now present 
	the linear program which tests for classical explainability (i.e., simplex-embeddability) of any set of quantum states and any set of quantum effects in terms of the matrices  $I_\Omega$, $I_\mathcal{E}$, $H_\Omega$, and $H_\mathcal{E}$, defined above and computed from the set of states and set of effects.
	
	\begin{lp}\label{thm:main}
		The Born rule statistics obtained by composing any state-effect  pair from $\Omega$ and $\mathcal{E}$ is classically-explainable if and only if the following linear program is satisfiable:
		\begin{subequations}
			\begin{align}
				\exists \ \sigma &\geq_e  0 \text{, an $m\times n$ matrix such that}\\\label{eq:QuantumCond}
				I_\mathcal{E}^T\cdot I_\Omega
				&= H_\mathcal{E}^T\cdot \sigma \cdot H_\Omega. 
		\end{align}\end{subequations}
	\end{lp}

	Note that if $\Omega$ and $\mathcal{E}$ span the full vector space of Hermitian operators, then the linear program simplifies somewhat, as the LHS of Eq.~\eqref{eq:QuantumCond} reduces to the identity map on $\mathsf{Herm}[\mathcal{H}]$.  Note that satisfiability is only a function of the cones defined by $\Omega$ and by $\mathcal{E}$, and so no other features of the states and effects are relevant to their nonclassicality, as  was also shown in Ref.~\cite{selby2021accessible,selby2021incompatibility}. A useful consequence of this fact is that $\Omega$ and $\mathcal{E}$ are classically-explainable if and only if their convex hulls are also classically-explainable.

	Testing for the existence of such a $\sigma$ is a linear program. In the repository~\cite{ourgithubrepo}, we give open-source Mathematica code for computing the relevant preliminaries and solving this linear program. The input to the code is simply a set of density matrices and a set of POVM elements (or, more generally, GPT state and effect vectors).  In practice the code runs in a few seconds for values of $n$ and $m$ up to around $20$.

	In the case that a classical explanation does exist, the code will output a specification of  an ontological model which represents the operational scenario in a noncontextual manner.
	This model can be computed from the matrix $\sigma$, as described in the Supplemental Material. In particular, every density matrix in $\rho \in \Omega$ is represented in the ontological model by a probability distribution $\mu_\rho$ over some set of ontic states $\Lambda$, while every POVM element in $\sigma \in \mathcal{E}$ is represented by a response function $\xi_\sigma$---that is, a $[0,1]-$valued function over $\Lambda$.
	Specifically, we compute a particular  non-negative factorization $\sigma = \beta \cdot \alpha$ where $\alpha:\mathds{R}^n \to \mathds{R}^\Lambda \geq_e 0$ and $\beta:\mathds{R}^\Lambda \to \mathds{R}^m \geq_e 0$, and then construct  linear maps $\tau_\Omega = \alpha \cdot H_\Omega$ and $\tau_\mathcal{E}:= \beta^T\cdot H_\mathcal{E}$, and use these to define  the epistemic states and response functions via 
	\beq
	\mu_\rho(\lambda):= [
	\tau_\Omega(\rho)]_\lambda \quad \text{and} \quad \xi_\sigma(\lambda):= [
	\tau_\mathcal{E}(\sigma)]_\lambda
	\eeq 
	for all $\lambda \in \Lambda$. That these functions are all non-negative follows from the definition of $H_\Omega$ and $H_\mathcal{E}$ together with element-wise non-negativity of $\alpha$ and $\beta$; that they are suitably normalised follows from the manner in which the decomposition into $\alpha$ and $\beta$ is chosen.  In particular,  the decomposition is constructed by taking $\beta= \sigma\cdot R$ and $\alpha = R^{-1}$ where $R$ is a diagonal rescaling matrix which ensures that $\xi_\mathds{1}(\lambda)=1$ for all $\lambda \in \Lambda$ (see Supplemental Material, Section C.I for details).
	Note that other choices for the decomposition of $\sigma= \beta\cdot \alpha$ are possible, and that this non-uniqueness translates into a non-uniqueness of the ontological model. 
	
	In the case that no solution exists, one can ask how much depolarising noise must be added to one's experiment until a solution becomes possible. This constitutes an operational measure of nonclassicality which we refer to as the {\em robustness of nonclassicality}.   Finding the minimal amount $r$ of noise is also a linear program:
	\begin{lp}\label{lpdepolarizing}
		Let $r$ be the minimum depolarising noise that must be added in order for the statistics obtained by 
		composing any state-effect pair from $\Omega$ and $\mathcal{E}$ to be classically-explainable. It can be computed by the linear program:
		\begin{subequations}
			\begin{align} \nonumber
				\text{minimise} \  \ &r \quad \text{such that} \\
				& \exists \ \sigma \geq_e  0 \text{, an $m\times n$ matrix such that}\\\label{eq:QuantumCondDep}
				& rI_\mathcal{E}^T\cdot D \cdot I_\Omega + (1-r)I_\mathcal{E}^T\cdot I_\Omega
				= H_\mathcal{E}^T\cdot \sigma \cdot H_\Omega, 
		\end{align}\end{subequations}
		where $D$ is the completely depolarising channel for the quantum system.
	\end{lp}
	Again, the corresponding ontological model can be straightforwardly computed from the matrix $\sigma$ found for the minimal value of $r$, and we give open-source code that returns both the value of $r$ and the associated model. 
	
	We also discuss in the Supplemental Material how one can easily adapt one's definition of robustness and the linear program for it to an arbitrary noise model.

	\noindent\emph{Examples---}
	Here we present three examples of sets of states and effects, and we assess the classical-explainability of their statistics using our linear program. In the case where the statistics are not classical, we also compute the noise robustness. A fully detailed analysis of these examples (including the explicit calculation of the matrices $H_\Omega$,  $H_\mathcal{E}$, $I_\Omega$, and $I_\mathcal{E}$), is given in the Supplemental Material. These specific examples are chosen to illustrate particular features of our approach, as we discuss therein.

	\noindent\textbf{Example 1:}
	Consider the set of four quantum states 
	\beq
	\Omega = \left\{ \ket{0}\bra{0} \,,\, \ket{1}\bra{1} \,,\, \ket{+}\bra{+} \,,\, \ket{-}\bra{-} \right\}\,
	\eeq
	on a qubit.
	In addition, consider the set of six effects 
	\beq\label{eq:Ex1Eff}
	\mathcal{E}=\left\{ \ket{0}\bra{0} \,,\, \ket{1}\bra{1} \,,\, \ket{+}\bra{+} \,,\, \ket{-}\bra{-} \,,\, \mathds{1}_2,0 \right\}\,.
	\eeq
	
	Next, consider the observable statistics---that is, the data that can be generated from any measurement constructed with these effects, when applied to any of these states.
	
	Our linear program finds that these statistics admit of a classical explanation.  This is to be expected, as this scenario is a subtheory of the noncontextual toy theory of Ref.~\cite{spekkens2007evidence} (namely, that given by restricting to the real plane). Indeed, this is the model which our code returns, and is depicted in Figure~\ref{fig:ome1}.

	\makebox[0pt][l]{%
		\hspace{-0.45cm}\begin{minipage}{0.48\textwidth}
			\begin{center}
				\noindent\[
				\begin{tikzpicture}[scale=1]
					\node at (-2,-4.4) {\footnotesize{(a) Embedding of states}};
					\node[draw,fill, color=gray, shape=circle,scale=.2] (a) at (0,0) {};
					\node[draw,fill, color=black, shape=circle,scale=.2] (b) at (-4,0) {};
					\node[draw,fill, color=black, shape=circle,scale=.2] (c) at (1,3) {};
					\node[draw,fill, color=black, shape=circle,scale=.2] (d) at (1,-3) {};
					\draw (b) -- (c) -- (d) -- (b) ; 
					\draw[dotted] (a) -- (b) ;
					\draw[dotted] (a) -- (c) ;
					\draw[dotted] (a) -- (d) ;
					\node[draw,fill, color=darkgreen, shape=circle,scale=.4] (z) at ($(a)!0.5!(c)$) {};
					\node[draw,fill, color=darkgreen, shape=circle,scale=.4] (p) at ($(c)!0.5!(d)$) {};
					\node[draw,fill, color=darkgreen, shape=circle,scale=.4] (o) at ($(d)!0.5!(b)$) {};
					\node[draw,fill, color=darkgreen, shape=circle,scale=.4] (m) at ($(a)!0.5!(b)$) {};
					\path[color=darkgreen, fill, fill opacity = 0.5] (z.center) -- (p.center) -- (o.center) -- (m.center) -- (z.center) -- cycle;
					\draw[thick, dashed, color=darkgreen] (o) -- (m) -- (z) -- (p) ;
					\draw[thick, color=darkgreen] (o) -- (p) ;
					\node [above left of = z, node distance=0.3cm] () {$\ket{0}$};
					\node [right of = p, node distance=0.35cm] () {$\ket{+}$};
					\node [below left of = o, node distance=0.3cm] () {$\ket{1}$};
					\node [above of = m, node distance=0.3cm] () {$\ket{-}$};
				\end{tikzpicture}
				\hspace{0.3cm}
				\begin{tikzpicture}[scale=0.8]
					
					\node[draw,fill, color=darkblue, shape=circle,scale=.4] (0101) at (-1.5,-0.5) {};
					\node[draw,fill, color=darkblue, shape=circle,scale=.4] (0011) at (0,0.5) {};
					\node[draw,fill, color=darkblue, shape=circle,scale=.4] (1100) at (2,-0.5) {};
					\node[draw,fill, color=darkblue, shape=circle,scale=.4] (1010) at (3.5,0.5) {};
					\node[draw,fill, color=darkblue, shape=circle,scale=.4] (1111) at (1,3.5) {};
					\node[draw,fill, color=darkblue, shape=circle,scale=.4] (0000) at (1,-3.5) {};
					
					\node [left of = 0101, node distance=0.35cm] () {$\ket{-}$};
					\node [right of = 1010, node distance=0.35cm] () {$\ket{+}$};
					\node [above right of = 0011, node distance=0.45cm] () {$\ket{1}$};
					\node [below left of = 1100, node distance=0.37cm] () {$\ket{0}$};
					\node [above of = 1111, node distance=0.3cm] () {$\mathds{1}$};
					\node [below of = 0000, node distance=0.3cm] () {$\mathbf{0}$};
					
					\draw[thick, color=darkblue] (1111) -- (0101) -- (0000) -- (1010) -- (1111) -- (1100) -- (0000) ;
					\draw[thick, color=darkblue] (0101) -- (1100) -- (1010) ;
					\draw[thick, dashed, color=darkblue] (1111) -- (0011) -- (0000) ;
					\draw[thick, dashed, color=darkblue] (0101) -- (0011) -- (1010) ;
					
					\path[color=darkblue, fill, fill opacity = 0.5] (1111.center) -- (1010.center) -- (0000.center) -- (0101.center) -- (1111.center) -- cycle;
					
					\node at (1,-5.5) {\footnotesize{(b) Embedding of effects}};
				\end{tikzpicture}
				\]
			\end{center}
			
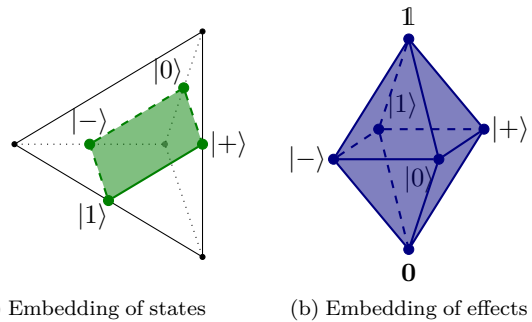
\captionof{figure}{\textbf{Classical explanation for Example 1} 
				(a) Depiction of the states in $\Omega$ (green dots), embedded in a 3-dimensional slice of a 4-dimensional simplex. (b) Depiction of the effects in $\mathcal{E}$ (blue dots), embedded in a 3-dimensional slice of the 4-dimensional hypercube that is dual to the simplex in (a).  Note that the convex hull of the effects happens to cover the entire hypercube in this particular slice.
				The simplex in (a) can be viewed as the set of probability distributions over a 4-element set $\Lambda$ of ontic states (black dots), while the hypercube in (b) can be viewed as the set of logically possible response functions for $\Lambda$. Hence, this simplex embedding corresponds to a noncontextual ontological model for---and hence~\cite{schmid2021characterization} a classical explanation of---the operational scenario. }
			\label{fig:ome1}
		\end{minipage}
	}

	\noindent\textbf{Example 2:}
	Consider the set of four quantum states 
	\beq
	\Omega = \{\ket{0}\bra{0}\,,\, \ket{1}\bra{1} \,,\, \ket{2}\bra{2}\,,\, \ket{3}\bra{3}\}\,
	\eeq
	on a four-dimensional quantum system.
	In addition, consider the set of six effects
	\begin{align}\label{sixeffects}
		\mathcal{E}=\{\ket{0}\bra{0}+\ket{1}\bra{1}\,,\,\ket{1}\bra{1}+\ket{2}\bra{2}\,,\,\ket{2}\bra{2}&+\ket{3}\bra{3}\,,\\ \nonumber
		&\ket{3}\bra{3}+\ket{0}\bra{0}\,,\, \mathds{1}_4\,,\, 0\}\,.  
	\end{align}
	
	Notably, the states and effects in this example do not span the same vector space. Still, our linear program also finds that the statistical data that arises from this admits of a classical explanation. This is a useful sanity check, since all the states and effects are diagonal in the same basis. We provide a depiction of the classical model which our code returns for this scenario in Figure~\ref{fig:ome2}.
	
	\makebox[0pt][l]{%
		\hspace{-0.45cm}\begin{minipage}{0.48\textwidth}
			\begin{center}
				\[
				\begin{tikzpicture}[scale=1]
					\node at (-2,-4.4) {\footnotesize{(a) Embedding of states}};
					\node[draw,fill, color=gray, shape=circle,scale=.4] (a) at (0,0) {};
					\node[draw,fill, color=darkgreen, shape=circle,scale=.4] (b) at (-4,0) {};
					\node[draw,fill, color=darkgreen, shape=circle,scale=.4] (c) at (1,3) {};
					\node[draw,fill, color=darkgreen, shape=circle,scale=.4] (d) at (1,-3) {};
					\draw[thick, color=darkgreen] (b) -- (c) -- (d) -- (b) ; 
					\draw[thick, dashed, color=darkgreen] (a) -- (b) ;
					\draw[thick, dashed, color=darkgreen] (a) -- (c) ;
					\draw[thick, dashed, color=darkgreen] (a) -- (d) ;
					\path[color=darkgreen, fill, fill opacity = 0.5] (b.center) -- (c.center) -- (d.center) -- cycle;
					\node [above left of = a, node distance=0.3cm] () {$\ket{0}$};
					\node [right of = c, node distance=0.35cm] () {$\ket{1}$};
					\node [right of = d, node distance=0.3cm] () {$\ket{2}$};
					\node [left of = b, node distance=0.3cm] () {$\ket{3}$};
				\end{tikzpicture}
				\hspace{0.1cm}
				\begin{tikzpicture}[scale=0.8]
					\node[draw,fill, color=darkblue, shape=circle,scale=.4] (0101) at (-1.5,-0.5) {};
					\node[draw,fill, color=darkblue, shape=circle,scale=.4] (0011) at (0,0.5) {};
					\node[draw,fill, color=darkblue, shape=circle,scale=.4] (1100) at (2,-0.5) {};
					\node[draw,fill, color=darkblue, shape=circle,scale=.4] (1010) at (3.5,0.5) {};
					\node[draw,fill, color=darkblue, shape=circle,scale=.4] (1111) at (1,3.5) {};
					\node[draw,fill, color=darkblue, shape=circle,scale=.4] (0000) at (1,-3.5) {};
					\node [left of = 0101, node distance=0.35cm] () {\textbf{E}${}_\mathbf{4}$};
					\node [right of = 1010, node distance=0.35cm] () {\textbf{E}${}_\mathbf{2}$};
					\node [above right of = 0011, node distance=0.45cm] () {\textbf{E}${}_\mathbf{1}$};
					\node [below left of = 1100, node distance=0.42cm] () {\textbf{E}${}_\mathbf{3}$};
					\node [above of = 1111, node distance=0.3cm] () {$\mathds{1}$};
					\node [below of = 0000, node distance=0.3cm] () {$\mathbf{0}$};
					\draw[thick, color=darkblue] (1111) -- (0101) -- (0000) -- (1010) -- (1111) -- (1100) -- (0000) ;
					\draw[thick, color=darkblue] (0101) -- (1100) -- (1010) ;
					\draw[thick, dashed, color=darkblue] (1111) -- (0011) -- (0000) ;
					\draw[thick, dashed, color=darkblue] (0101) -- (0011) -- (1010) ;
					\path[color=darkblue, fill, fill opacity = 0.5] (1111.center) -- (1010.center) -- (0000.center) -- (0101.center) -- (1111.center) -- cycle;
					\node at (1,-5.5) {\footnotesize{(b) Embedding of effects}};
				\end{tikzpicture} \]
			\end{center}
			\captionof{figure}{\textbf{Classical explanation for Example 2.} 
				(a) Depiction of the states in $\Omega$ (green dots), embedded in a 3-dimensional slice of a 4-dimensional simplex. 
				(b) Depiction of the effects in $\mathcal{E}$ (blue dots), embedded in a 3-dimensional slice of the 4-dimensional hypercube that is dual to the simplex in (a).
				Note that the convex hull of the states (effects) happens to cover the entire simplex (hypercube) in this particular slice.
				Exactly as in the last example, this simplex embedding corresponds to a noncontextual ontological model for---and hence~\cite{schmid2021characterization} a classical explanation of---the operational scenario.}\label{fig:ome2}
	\end{minipage}}

	\noindent\textbf{Example 3:} Our third example is obtained from the first example by rotating all of the effects by an angle of $\frac{\pi}{4}$ about the $\sigma_y$ axis.  (This is the set of states and effects relevant for parity-oblivious multiplexing~\cite{POM}.) 
	In this case, our linear program finds that there is no classical explanation of the observable statistics. Moreover, it finds that the depolarizing-noise robustness for these states and effects is $r=1 - \frac{1}{\sqrt{2}} \sim 0.3$. In Fig.~\ref{fig:ome3} we depict the classical model for the case of depolarisation at this noise threshold.
	
	\makebox[0pt][l]{%
		\hspace{-0.45cm}\begin{minipage}{0.48\textwidth}
			\begin{center}\[
				\begin{tikzpicture}[scale=1]
					\node at (-2,-4.4) {\footnotesize{(a) Embedding of states}};
					\node[draw,fill, color=gray, shape=circle,scale=.2] (a) at (0,0) {};
					\node[draw,fill, color=black, shape=circle,scale=.2] (b) at (-4,0) {};
					\node[draw,fill, color=black, shape=circle,scale=.2] (c) at (1,3) {};
					\node[draw,fill, color=black, shape=circle,scale=.2] (d) at (1,-3) {};
					\draw (b) -- (c) -- (d) -- (b) ; 
					\draw[dotted] (a) -- (b) ;
					\draw[dotted] (a) -- (c) ;
					\draw[dotted] (a) -- (d) ;
					\node[draw,fill, color=darkgreen, shape=circle,scale=.4] (z) at ($(a)!0.5!(c)$) {};
					\node[draw,fill, color=darkgreen, shape=circle,scale=.4] (p) at ($(c)!0.5!(d)$) {};
					\node[draw,fill, color=darkgreen, shape=circle,scale=.4] (o) at ($(d)!0.5!(b)$) {};
					\node[draw,fill, color=darkgreen, shape=circle,scale=.4] (m) at ($(a)!0.5!(b)$) {};
					\path[color=darkgreen, fill, fill opacity = 0.5] (z.center) -- (p.center) -- (o.center) -- (m.center) -- (z.center) -- cycle;
					\draw[thick, dashed, color=darkgreen] (o) -- (m) -- (z) -- (p) ;
					\draw[thick, color=darkgreen] (o) -- (p) ;
					\node [above left of = z, node distance=0.3cm] () {$\ket{0}$};
					\node [right of = p, node distance=0.35cm] () {$\ket{+}$};
					\node [below left of = o, node distance=0.3cm] () {$\ket{1}$};
					\node [above of = m, node distance=0.3cm] () {$\ket{-}$};
				\end{tikzpicture}
				\hskip 0.5cm
				\begin{tikzpicture}[scale=0.8]
					
					\node at (1,-5.5) {\footnotesize{(b) Embedding of effects}};
					
					\node[] (0101) at (-1.5,-0.5) {};
					\node[] (0011) at (0,0.5) {};
					\node[] (1100) at (2,-0.5) {};
					\node[] (1010) at (3.5,0.5) {};
					\node[draw,fill, color=darkblue, shape=circle,scale=.4] (1111) at (1,3.5) {};
					\node[draw,fill, color=darkblue, shape=circle,scale=.4] (0000) at (1,-3.5) {};
					
					\node [above of = 1111, node distance=0.3cm] () {$\mathds{1}$};
					\node [below of = 0000, node distance=0.3cm] () {$\mathbf{0}$};
					
					\draw (1111) --   (1100.center) -- (0000) -- (0101.center) -- (1111) -- (1010.center) -- (0000);
					\draw[thick] (0101.center) -- (1100.center) -- (1010.center) ;
					
					\draw[thick, dotted] (1111) -- (0011.center) -- (0000) ;
					\draw[thick, dotted] (0101.center) -- (0011.center) -- (1010.center) ;

					\node [draw,fill, color=darkblue, shape=circle,scale=.4] (m1) at ($(1010)!0.5!(1100)$) {};
					\node [draw,fill, color=darkblue, shape=circle,scale=.4] (m2) at ($(1100)!0.5!(0101)$) {};
					\node [draw,fill, color=darkblue, shape=circle,scale=.4] (m3) at ($(0101)!0.5!(0011)$) {};
					\node [draw,fill, color=darkblue, shape=circle,scale=.4] (m4) at ($(0011)!0.5!(1010)$) {};
					
					\draw[thick, color=darkblue]  (1111.center) -- (m2) -- (0000.center) -- (m1) -- (1111.center);
					
					\draw[thick, dashed, color=darkblue]  (1111.center) -- (m4) -- (0000.center) -- (m3) -- (1111.center);
					
					\draw[thick, dashed, color=darkblue]  (m1) -- (m2) -- (m3) -- (m4) -- (m1);
					
					\node [right of = m1, node distance=0.3cm] () {$\mathbf{F}_1$};
					\node [below left of = m2, node distance=0.35cm] () {$\mathbf{F}_2$};
					\node [below left of = m3, node distance=0.35cm] () {$\mathbf{F}_3$};
					\node [above left of = m4, node distance=0.35cm] () {$\mathbf{F}_4$};
					
					\path[color=darkblue, fill, fill opacity = 0.5] (1111.center) -- (m1.center) -- (0000.center) -- (m3.center) -- (1111.center) -- cycle;
					
				\end{tikzpicture}\]
			\end{center}
			
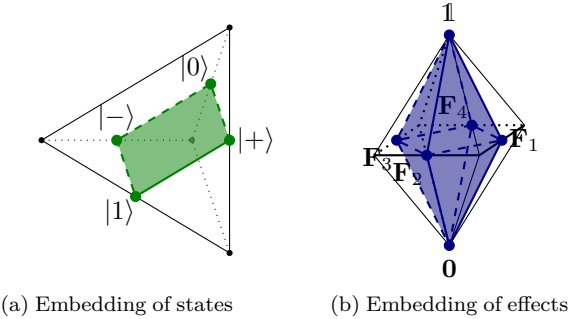
\captionof{figure}{\textbf{Classical explanation for Example 3 when depolarised by $r=1 - \frac{1}{\sqrt{2}}$.} 
				(a) Depiction of the states in $\Omega$ (green dots), embedded in a 2-dimensional slice of a 4-dimensional simplex. 
				(b) Depiction of the effects in $\mathcal{E}$ (blue dots), embedded in a 3-dimensional slice of the 4-dimensional hypercube that is dual to the simplex in (a).
				Exactly as in the last example, this simplex embedding corresponds to a noncontextual ontological model for---and hence~\cite{schmid2021characterization} a classical explanation of---the depolarised operational scenario. If the depolarisation was less strong, then such a noncontextual ontological model would not exist. Visually, we can get some intuition for this by observing that if we grow either the green square or the red octahedron, then we would end up with the states or effects lying outside of the simplex or hypercube.}\label{fig:ome3}
	\end{minipage}}

	\noindent\emph{Related  linear programs---}
	We reiterate that the core of our  linear program is closely related to the linear program introduced in Section~4.2 of Ref.~\cite{gitton2020solvable} as specialised to the polytopic case (that we consider here) in Section~4.3 of Ref.~\cite{gitton2020solvable}. However, the approach of Ref.~\cite{gitton2020solvable} differs from ours in a critical preprocessing step, and so its assessment of classicality differs from ours in some examples. Indeed, their proposal deems Example 2 nonclassical, while our approach deems it classical. But, the `nonclassical' verdict is clearly mistaken, since all the states and effects in that example are simultaneously diagonalizable. Still, we emphasize that the mathematical tools of Ref.~\cite{gitton2020solvable} are quite useful and applicable to our notion of classicality, and indeed even extend some results to non-polytopic GPTs (although in this case, testing for nonclassicality is likely not a linear program) via inner and outer polytopic approximations as discussed in Section~4.4 of Ref.~\cite{gitton2020solvable}. 
	
	Ref.~\cite{Schmid2018} also presented a linear programming approach which could determine if a prepare-measure scenario admits of a noncontextual model or not. In that work, however, the input to the linear program required the specification of a set of operational equivalences for the states and  another set for the effects. In contrast, in the current work, the input to the algorithm is simply a set of quantum (or GPT) states and effects. 
	The full set of operational equivalences that hold among states and among effects are  derivable from this input; however, one need not consider them explicitly. The linear program we present here determines if there is a noncontextual model with respect to {\em all} of the operational equivalences that hold in quantum theory (or within the given GPT). 
	
	Ref.~\cite{merminsquare} provided another linear programming approach to testing noncontextuality in the context of a particular class of prepare-measure scenarios, namely those wherein all operational equivalences arise  from different ensembles of preparation procedures, all of which define the same average state. Using the flag-convexification technique of Ref.~\cite{selby2021incompatibility,selby2021accessible}, we suspect that {\em all} prepare-measure scenarios can be transformed into prepare-measure scenarios of this particular type, in which case the linear program from Ref.~\cite{merminsquare} would be as general as the approach we have discussed herein. However, this remains to be proven.
	
	An interesting open question is to determine the relative efficiency of these algorithms.

	\noindent\emph{Closing remarks---}
	Our arguments in the Supplemental Material demonstrate that if a noncontextual model exists for a scenario, then there also exists a model  with $\mathsf{dim}[S_\Omega]\mathsf{dim}[S_\mathcal{E}] \leq \mathsf{dim}[\mathcal{H}]^2$ ontic states (or less),  
	This bound was first proven in Ref.~\cite{gitton2020solvable} by similar arguments.
	It is not yet clear if this bound is tight.
	
	Additionally, our arguments in the Supplemental Material hinge on the existence of a particular kind of decomposition of the identity channel. The arguments proving a structure theorem for noncontextual models in Ref.~\cite{schmid2020structure} hinged on a similar decomposition of the identity channel, and it would be interesting to investigate this connection further.
	We hope that a synthesis of the algorithmic techniques herein with the compositional techniques of Refs.~\cite{schmid2020structure,schmid2020unscrambling} might lead to algorithms for testing nonclassicality in prepare-transform-measure scenarios and eventually in arbitrary circuits.
	
	In Ref.~\cite{mueller2021testing}, the definition of simplex-embedding was generalized to embeddings into arbitrary GPTs. It would be interesting to investigate whether similar programs (albeit most likely not linear ones\footnote{E.g., in the quantum case these are typically SDP hierarchies \cite{Chaturvedi2021characterising}}.) could be developed for testing for such embeddings.
	
	Finally, we note that our linear program  and open source implementation are 
	ideally suited for proving nonclassicality in real experiments~\cite{mazurek2016experimental}, especially when coupled with theory-agnostic tomography techniques~\cite{AgnosticTomogr,mazurek2017experimentally}.

	\noindent\emph{Acknowledgements---}
	We acknowledge useful feedback from Rob Spekkens.  JHS was supported by the National Science Centre, Poland (Opus project, Categorical Foundations of the Non-Classicality of Nature, project no. 2021/41/B/ST2/03149). ABS, DS, and VPR were supported by the Foundation for Polish Science (IRAP project, ICTQT, contract no. MAB/2018/5, co-financed by EU within Smart Growth Operational Programme). EW was supported by Perimeter Institute for Theoretical
	Physics. Research at Perimeter Institute is supported
	in part by the Government of Canada through the Department of Innovation, Science and Economic Development Canada and by the Province of Ontario through
	the Ministry of Colleges and Universities.
	All diagrams were made using TikZiT.
 
\bibliographystyle{apsrev4-2-wolfe}
\setlength{\bibsep}{3pt plus 3pt minus 2pt}
\nocite{apsrev42Control}
\bibliography{bib}

\renewcommand\appendixname{Supplemental Material}

\newpage

\newpage

\appendix

\addcontentsline{toc}{section}{Appendix} 
\part{Appendix} 
\parttoc 

\addcontentsline{toc}{section}{\quad \hspace*{\fill} \quad }

	\section{Related work}\label{se:comparisonothers}
	
	The examples we examined in Section~\ref{ap:workedex} are useful for comparing generalized noncontextuality with two other purported notions of classicality, introduced by Shahandeh in Ref.~\cite{Shahandeh2017} and by Gitton and Woods in Ref.~\cite{gitton2020solvable}.  After this, we also comment on a third related  idea  from Ref.~\cite{mueller2021testing}.

	\subsection{Shahandeh's  proposed  notion of classicality}

	In Ref.~\cite{shahandeh2021contextuality}, Shahandeh proposes that an operational theory whose GPT has dimension $d$ should be deemed classically-explainable if and only if it admits of a noncontextual model {\em whose ontic state space contains exactly $d$ ontic states}. Clearly, this implies generalized noncontextuality, but it is a strictly more stringent condition. This can be seen  from the fact that Example 1 from the  main text admits of a generalized noncontextual model, and noting that this model {\em necessarily} requires 4 ontic states~\cite{schmid2021characterization}, while $d=3$. Our technique deems this example classically-explainable (and provides a noncontextual model for it), while the example is considered not classically-explainable by Shahandeh's definition.
	
	Despite Shahandeh's arguments in Ref.~\cite{shahandeh2021contextuality} for including this dimensional restriction (as well as our own attempts to find a motivation for it, e.g., by appealing to some variant of Leibniz's principle), we have not yet seen any compelling motivations for it.

	\subsection{Gitton-Woods's  proposed  notion of classicality}
	
	In Ref.~\cite{gitton2020solvable}, Gitton and Woods propose a novel notion of nonclassicality. To understand their proposal, first recall that generalized noncontextuality only implies constraints on the ontological representations of operational processes which are genuinely operationally equivalent---that is, those that give exactly the same predictions for any operational scenario in which they appear. For example, two preparation procedures are only operationally equivalent if they give the same statistics for the outcomes {\em of all possible measurements}, and it is only in this case that one has reason to expect that their ontological representations be identical as stochastic processes (as the principle of noncontextuality demands). 
	
	In contrast, Ref.~\cite{gitton2020solvable} proposes that we drop this requirement, and instead demand that operational processes which make the same predictions {\em for the particular measurements performed in a given experiment} should {\em also} have the same ontological representations as stochastic processes, even if these operational processes are ultimately distinguishable by other procedures (not considered in the given experiment). 
	As an example, this approach would demand that the ontological representations of 
	$|0\rangle + i |1\rangle$ and $|0\rangle - i |1\rangle$ 
	must be identical in any experiment wherein the only measurements performed on the system were confined to the $X-Z$ plane, since all such measurements give the same statistics on these two states.
	
	The notion of classicality that results in Ref.~\cite{gitton2020solvable} is hence distinct from generalized noncontextuality. One can see this explicitly from a particular example:  it deems our Example 2 nonclassical\footnote{To see that it is deemed nonclassical by the approach of Ref.~\cite{gitton2020solvable}, it suffices to note that if one quotients its (three-dimensional) realized state space with respect to its (two-dimensional) realized effect space, the resulting GPT is Boxworld, as in our  fourth example, which we show is not simplex-embeddable.}, even though there exists a generalized noncontextual ontological model for it, as we demonstrated in Section~\ref{ap:workedex}. 
	
	Recall, however, that our Example 2 involved only states and effects which are diagonal in a fixed basis. Given that simultaneously diagonalizability is  widely considered to be a strong notion of classicality, it is clear that Gitton-Woods' proposal is problematic. 
	
	In addition, their approach does not respect the well motivated constraint that any subset of states and effects from a classically-explainable theory must also be classically-explainable. (One can verify this by considering our Example 2 as a sub-fragment of the full set of quantum states and effects in the computational basis of a 4-dimensional Hilbert space, and noting that Gitton-Woods' proposal deems this strictly larger scenario classical\footnote{This can be seen from the fact that their approach coincides with ours when the states and effects span the same space.}).

	These examples illustrate the fact that one's assessment of nonclassicality, at least in the sense of generalised noncontextuality, is only as good as one's knowledge of the true operational equivalences. If one quotients with respect to a set of operational procedures which is {\em not} tomographically complete for the system under consideration, then one can mistakenly conclude that one's experiment does not admit of a classical explanation. For more on this, see the discussions of tomographic completeness in Refs.~\cite{gencontext,pusey2019contextuality,schmid2023addressing}, and see also Refs.~\cite{mazurek2017experimentally,AgnosticTomogr} for methods for obtaining evidence that one truly has tomographically complete ways of probing a given system. 
	
	The mathematical tools developed by Gitton and Woods can be reconsidered as {\em sufficient} (but not necessary) conditions for classical-explainability (defined by existence of a generalized noncontextual model).  Moreover,  other elements of the analysis in Ref.~\cite{gitton2020solvable} are quite useful; as we noted above, for example, Ref.~\cite{gitton2020solvable} essentially derives the same linear program that we developed (independently) and presented here  and provides useful techniques for polytopic approximations to situations involving infinite sets of states and effects.  
	
	\subsection{M{\"u}ller-Garner's notion of simulability} 
	
	Finally, we comment on some related ideas from Ref.~\cite{mueller2021testing}, in which M{\"u}ller and Garner define a notion of `classical simulation'. A simulation associates to each state (effect) of a given GPT a {\em set} of states (effects) in a simplicial GPT, such that a few natural conditions hold (e.g., probabilities are reproduced, and convex mixture is respected in the appropriate sense). 
	
	M{\"u}ller and Garner do not claim that this is a notion of classicality; indeed, they reprove a result by Holevo establishing that every GPT has a classical `simulation' of this sort. Hence, this notion of simulability cannot form the basis of a   useful  notion of classicality, as it does not establish a meaningful dividing line between classical and nonclassical phenomena.
	
	As an aside, it is interesting to note that the mathematical object which simulates a given GPT can be viewed as an epistemically restricted ontological theory~\cite{spekkens2007evidence,epistricted,bartlett2012reconstruction,catani2021interference}. This is analogous to how any simplex-embeddable GPT can be viewed as an epistemically restricted ontological theory. However, in the latter case, the epistemic states and response functions allowed by the epistemic restriction necessarily span the same vector space\footnote{One can verify that this property is satisfied in all of the epistemically restricted theories studied to date~\cite{spekkens2007evidence,epistricted}, as required by the fact that they constitute noncontextual models. }, whereas in the former, this need not be the case. 
	
	Ontological theories which are epistemically restricted in this manner (where the epistemic states and response functions do not span the same vector space) are consequently able to reproduce {\em all} operational theories, including those which do not admit of any noncontextual representation. An interesting question for future work would be to provide independent physical motivations for rejecting epistemic restrictions of this sort as natural classical explanations. 
	
	\section{Formal definitions}

	\subsection{Generalized probabilistic theories}
	
	In the main text, we presented our results  using the formalism of quantum theory.
	In the following, we state and prove our results in a more general manner which does not assume the validity of quantum theory. We do this within the framework of generalized probabilistic theories (GPTs). 
	
	Ref.~\cite{selby2021accessible} contains a comprehensive introduction to the particular formalization of GPTs which we use here. In addition, see Refs.~\cite{muller2021probabilistic,lami2018non,plavala2021general} for reviews of GPTs more broadly, and Refs.~\cite{chiribella2016quantum,hardy2011reformulating} for the diagrammatic formalism for GPTs that we use here. In brief, a GPT describes a possible theory of the world, as characterized by its operational statistics~\cite{Hardy,GPT_Barrett}. By ranging over different GPTs, then, one ranges over a landscape of possible ways the world might be. 
	
	We now briefly review
	the GPT description of prepare-measure scenarios, which are the focus of our manuscript.
	In this context, a GPT is formally defined by a quadruple 
	\beq
	\mathcal{G}=\left(\left\{%
\InputIfFileExists{Diagrams/GPTState.tikz}{}{\input{./figures/Diagrams/GPTState.tikz}} \right\}_{ s\in \Omega^\mathcal{G}},\  \left\{%
\InputIfFileExists{Diagrams/GPTEffect.tikz}{}{\input{./figures/Diagrams/GPTEffect.tikz}}\right\}_{ e\in \mathcal{E}^\mathcal{G}},\  %
\InputIfFileExists{Diagrams/ident.tikz}{}{\input{./figures/Diagrams/ident.tikz}} ,\ %
\InputIfFileExists{Diagrams/discardGPT.tikz}{}{\input{./figures/Diagrams/discardGPT.tikz}} \right) \,,\eeq
	where $\Omega^\mathcal{G}$ is a convex set of GPT states which span a real vector space $S$, and $\mathcal{E}^\mathcal{G}$ is a convex set of GPT effects which span the dual space $S^*$. 
	Each GPT state represents an operational preparation procedure---possibly one that occurs with non-unit probability\footnote{That is, we take $\Omega^\mathcal{G}$ to include subnormalised states, representing preparation procedures which fail with some nonzero probability, as this will be convenient later in the paper.}, and each GPT effect represents an operational measurement procedure together with the observation of a particular outcome.
	We follow the standard convention of assuming that the sets $\Omega^\mathcal{G}$ and $\mathcal{E}^\mathcal{G}$ are  finite-dimensional, convex, and compact. The quadruple also specifies a probability rule (via the identity map $\mathds{1}_S$---see Eq.~\eqref{eq:GPTProbRule}), and a unit effect $\discard\in\mathcal{E}^\mathcal{G}$. These are in fact both redundant in the case of standard GPTs, but we have included them here to highlight the fact that standard GPTs are special cases of accessible GPT fragments, a concept  we introduced in Ref.~\cite{selby2021accessible} and which we review  in the next section.

	A measurement in a GPT is a set of effects (one for each possible outcome) summing to the privileged unit effect, $\discard$. 
	In any measurement containing a GPT effect $e\in\mathcal{E}^\mathcal{G}$, the probability of the outcome corresponding to that effect arising, given a preparation of the system in a state described by the GPT state $s\in\Omega^\mathcal{G}$, is given by
	\beq\label{eq:GPTProbRule}
	\mathrm{Prob}\left(%
\InputIfFileExists{Diagrams/GPTEffect.tikz}{}{\input{./figures/Diagrams/GPTEffect.tikz}},\ %
\InputIfFileExists{Diagrams/GPTState.tikz}{}{\input{./figures/Diagrams/GPTState.tikz}}\right)\quad :=\quad %
\InputIfFileExists{Diagrams/GPTPRule.tikz}{}{\input{./figures/Diagrams/GPTPRule.tikz}}.
	\eeq
	
	The set of states and effects in any valid GPT must satisfy a number of constraints, three of which we highlight here:
	\ben
	\item The principle of {\em tomography}~\cite{Hardy,GPT_Barrett} must be satisfied.
	This means that all states and effects can be uniquely identified by the predictions they generate. Formally, for GPT states it means that $e(s_1)=e(s_2)$ for all $e\in \mathcal{E}^\mathcal{G}$ if and only if $s_1=s_2$; for GPT effects, it means that $e_1(s)=e_2(s)$ for all $s\in \Omega^\mathcal{G}$ if and only if $e_1=e_2$. 
	\item For every state $s\in \Omega^\mathcal{G}$, it holds that 
	$\frac{1}{\discard(s)} s \in \Omega^\mathcal{G}$.
	That is, for every state in the GPT, the normalised counterpart is also in the GPT.
	This is an important constraint first  highlighted in Ref.~\cite{chiribella2010probabilistic}.
	\item For all $e\in\mathcal{E}^\mathcal{G}$, there exists $e^\perp\in \mathcal{E}^\mathcal{G}$ such that $e+e^\perp=\discard$.
	\een
	We highlight these in particular as they are relevant for this manuscript. Notably, we will relax the first two conditions in the following section.
	
	In this paper we are interested in whether or not a given GPT, or an experiment performed within a given GPT, is classically explainable. In Refs.~\cite{schmid2021characterization,selby2021accessible}, it was shown that the appropriate notion of classical-explainability is the notion of simplex-embeddability.
	This geometric criterion deems a GPT \textit{classically-explainable} if its state space can be embedded into a simplex (of any dimension) and its effect space can be embedded into the dual to the simplex, such that probabilities are preserved. This notion is motivated by the fact that the existence of a simplex embedding for a GPT is equivalent to the existence of a generalized noncontextual ontological model for any operational scenario which leads (through quotienting by operational equivalences) to the GPT.  In turn, recall (e.g., from our introduction) that there are many motivations for taking generalized noncontextuality as one's notion of classical explainability for operational scenarios. Simplex-embeddability can also be motivated as a notion of classical-explainability by the independent consideration that simplicial GPTs are the standard way of capturing strictly classical theories---i.e., those wherein all possible measurements are compatible.
	We refer the reader to Refs.~\cite{schmid2021characterization,selby2021accessible} for more details on this and on the closely related notion of simplicial-cone embedding.

	\begin{definition}[Simplicial-cone embeddings and simplex embeddings of a GPT] \label{def:SimplexEmbeddingGPT}
		A \emph{simplicial-cone embedding}, $\tau_\mathcal{G}$, of a GPT, $\mathcal{G}$,
		is defined by a set of ontic states $\Lambda$ and a pair of linear maps
		\beq
\InputIfFileExists{Diagrams/iotaGPT.tikz}{}{\input{./figures/Diagrams/iotaGPT.tikz}} \quad\text{and}\quad %
\InputIfFileExists{Diagrams/kappaGPT.tikz}{}{\input{./figures/Diagrams/kappaGPT.tikz}}
		\eeq
		such that for all $s\in \Omega^\mathcal{G}$ and for all $e\in \mathcal{E}^\mathcal{G}$ we have
		\beq
\InputIfFileExists{Diagrams/IotaStateGPT.tikz}{}{\input{./figures/Diagrams/IotaStateGPT.tikz}} \ \geq_e\  0\quad\text{,}\quad %
\InputIfFileExists{Diagrams/KappaEffectGPT.tikz}{}{\input{./figures/Diagrams/KappaEffectGPT.tikz}} \ \geq_e\  0
		\eeq
		and such that
		\beq\label{eq:SEProbGPT}
\InputIfFileExists{Diagrams/probRuleGPT.tikz}{}{\input{./figures/Diagrams/probRuleGPT.tikz}}\quad=\quad %
\InputIfFileExists{Diagrams/probRuleSimplexDecompGPT.tikz}{}{\input{./figures/Diagrams/probRuleSimplexDecompGPT.tikz}}.
		\eeq
		
		A simplicial-cone embedding is said to be a \emph{simplex embedding} if it moreover satisfies
		\beq
\InputIfFileExists{Diagrams/kappaCausalGPT.tikz}{}{\input{./figures/Diagrams/kappaCausalGPT.tikz}}\quad=\quad%
\InputIfFileExists{Diagrams/kappaCausal1.tikz}{}{\input{./figures/Diagrams/kappaCausal1.tikz}}.
		\eeq
	\end{definition}
	Although a simplex embedding must satisfy this additional constraint, we proved in Ref.~\cite{selby2021accessible} that a simplicial-cone embedding exists if and only if a simplex embedding exists. We expand on this in Section~\ref{app:NCOM}. 
	
	We note also that a simplex-embedding of a GPT is equivalent to an ontological model of a GPT~\cite{schmid2021characterization,schmid2020structure}. It follows, then, that an operational theory (or scenario or experiment) admits of a noncontextual ontological model if and only if the associated GPT (or GPT fragment,  see next subsection) admits of an ontological model.
	
	Here and throughout, we will denote physical processes in white, and mathematical processes (like embedding, projection, and inclusion maps) in black.
	
	Note that since states are spanning for $S$ and effects are spanning for $S^*$, we can equivalently write Eq.~\eqref{eq:SEProbGPT}, which expresses the constraint that the operational data is reproduced by the embedding, as
	\beq\label{eq:SEProbGPTEquiv1}
\InputIfFileExists{Diagrams/probRuleGPTEquiv.tikz}{}{\input{./figures/Diagrams/probRuleGPTEquiv.tikz}}\quad=\quad %
\InputIfFileExists{Diagrams/probRuleSimplexDecompGPTEquiv.tikz}{}{\input{./figures/Diagrams/probRuleSimplexDecompGPTEquiv.tikz}}.
	\eeq
	
	\subsection{GPT fragments}

	In a standard GPT (as defined in the previous section), {\em all} states and effects which are taken to be physically possible given one's theory of the world are required to be included in the sets $\Omega^\mathcal{G}$ and $\mathcal{E}^\mathcal{G}$. When applying the framework of GPTs to describe {\em particular experiments} rather than possible theories of the world, however, one must drop this requirement.

	From a practical perspective, a specific prepare-measure experiment can be described simply as a subset $\Omega^\mathfrak{F} \subseteq \Omega^\mathcal{G}$
	representing the preparation procedures in the experiment and a subset $\mathcal{E}^\mathfrak{F} \subseteq \mathcal{E}^\mathcal{G}$ of effects representing the measurement outcomes in the experiment.  We will refer to this object as a {\em GPT fragment}. 
	More formally:
	\begin{definition}[GPT fragment]\label{gptfragdefn}
		A GPT fragment, $\mathfrak{F}$, is specified by the underlying GPT, $\mathcal{G}$, together with a designated subset of states $\Omega^\mathfrak{F}  \subseteq \Omega^\mathcal{G}$ and of effects $\mathcal{E}^\mathfrak{F} \subseteq \mathcal{E}^\mathcal{G}$. \end{definition}
	
	Critically, the sets of states and effects in a GPT fragment ($\Omega^\mathfrak{F}$ and $\mathcal{E}^\mathfrak{F}$) need not satisfy all the constraints that a GPT must satisfy. In particular, 
	(i) the set of state vectors and effect covectors in a GPT fragment need not be tomographically complete for each other (i.e., they need not span the same vector space and its dual,  respectively), and
	(ii) the set of state vectors in a GPT fragment may contain subnormalized states whose normalised counterparts are not in the GPT fragment.

	One can then ask whether the fragment (as opposed to the underlying GPT) is classically-explainable. The appropriate notion of classical-explainability for GPT fragments follows immediately from the notion for the underlying GPT, namely Definition~\ref{def:SimplexEmbeddingGPT}:
	\begin{definition}[Simplicial-cone embeddings and simplex embeddings of GPT fragments]\label{def:SEofGPTF}
		Definition~\ref{def:SimplexEmbeddingGPT}, but where one replaces $\Omega^\mathcal{G}$ with $\Omega^\mathfrak{F}$ and replaces $\mathcal{E}^\mathcal{G}$ with $\mathcal{E}^\mathfrak{F}$. 
	\end{definition}
	Note that any fragment $\mathfrak{F}$ of a classically-explainable underlying GPT $\mathcal{G}$ is necessarily also classically-explainable; contrapositively, if a fragment is not classically-explainable, then neither is the underlying GPT.

	Note that $\Omega^\mathfrak{F}$ and $\mathcal{E}^\mathfrak{F}$ are not necessarily spanning for the underlying GPT vector space $S$  and dual space $S^*$, respectively.  As such, one can no longer derive Eq.~\eqref{eq:SEProbGPTEquiv1} as an equivalent way to capture the constraint that the operational predictions be reproduced (as in Eq.~\eqref{eq:SEProbGPT}). However, we can derive an analogous condition by introducing some projection maps. Although this is not necessary at this stage, these maps will be useful in the next section.
	
	Define a particular pair of idempotent linear maps 
	\beq %
\InputIfFileExists{Diagrams/StateProjection.tikz}{}{\input{./figures/Diagrams/StateProjection.tikz}}\quad\text{and}\quad %
\InputIfFileExists{Diagrams/EffectProjection.tikz}{}{\input{./figures/Diagrams/EffectProjection.tikz}},\eeq
	where idempotence means that
	\beq
\InputIfFileExists{Diagrams/StateProjection.tikz}{}{\input{./figures/Diagrams/StateProjection.tikz}}\ \ =\ \  %
\InputIfFileExists{Diagrams/StateProjectionIdempotent.tikz}{}{\input{./figures/Diagrams/StateProjectionIdempotent.tikz}}\quad\text{and}\quad %
\InputIfFileExists{Diagrams/EffectProjection.tikz}{}{\input{./figures/Diagrams/EffectProjection.tikz}} \ \ =\ \  %
\InputIfFileExists{Diagrams/EffectProjectionIdempotent.tikz}{}{\input{./figures/Diagrams/EffectProjectionIdempotent.tikz}}.
	\eeq
	The defining feature of these idempotents is that they characterize the subspaces of states and effects in the fragment via
	\beq
\InputIfFileExists{Diagrams/ProjectedState.tikz}{}{\input{./figures/Diagrams/ProjectedState.tikz}}\ \  =\ \  %
\InputIfFileExists{Diagrams/State.tikz}{}{\input{./figures/Diagrams/State.tikz}} \quad \iff \quad s \in \mathsf{Span}[\Omega^\mathfrak{F}]
	\eeq
	and 
	\beq
\InputIfFileExists{Diagrams/ProjectedEffect.tikz}{}{\input{./figures/Diagrams/ProjectedEffect.tikz}}\ \  =\ \  %
\InputIfFileExists{Diagrams/Effect.tikz}{}{\input{./figures/Diagrams/Effect.tikz}} \quad \iff \quad e \in \mathsf{Span}[\mathcal{E}^\mathfrak{F}].
	\eeq
	Although more than one idempotent map may satisfy these constraints, they are related by reversible linear maps relating two different choices of bases for $S$, and so we will see that the results hold for any choice satisfying these conditions.
	Then, Eq.~\eqref{eq:SEProbGPTEquiv1} can be more generally expressed as
	\beq\label{eq:FragmentPRuleEquiv}
\InputIfFileExists{Diagrams/FragmentPRule1.tikz}{}{\input{./figures/Diagrams/FragmentPRule1.tikz}}\quad =\quad %
\InputIfFileExists{Diagrams/FragmentPRule2.tikz}{}{\input{./figures/Diagrams/FragmentPRule2.tikz}},
	\eeq
	which now directly applies to GPT fragments as well. These idempotents and this equivalent characterisation of Eq.~\eqref{eq:SEProbGPT} will be useful in the following section. 
	
	\subsection{Accessible GPT fragments}\label{sec:AccGPTFrag}
	
	As one can see from Definition~\ref{gptfragdefn}, a GPT fragment is explicitly defined {\em with respect} to an underlying GPT. 
	
	From a theorist's perspective, however, it can be convenient to work with an `intrinsic' characterisation of an experiment, rather than viewing it as a fragment living inside an underlying GPT. That is, it is often useful to view one's subsets of states and effects as living in the vector spaces which they span, rather than the vector space of the underlying GPT (a vector space that will generally be of larger dimension).  This resulting object has been termed an {\em accessible GPT fragment}~\cite{selby2021accessible}.  
	
	The definition of an accessible GPT fragment in Ref.~\cite{selby2021accessible}  also incorporates a closure of the state space and of the effect space under classical processings---convex mixtures, coarse-grainings of outcomes, and so on.\footnote{Note that the choice to incorporate this closure is in some sense optional; one could also study objects like accessible GPT fragments, but without this closure.} Thus, they represent all states and effects that are {\em accessible} given the laboratory devices in question (but, like with GPT fragments, they need not represent all states and effects that are physically possible in the underlying theory).
	As a consequence of this closure under classical processings, all accessible GPT fragments share some geometric structure; however, this additional structure is not needed for proving our results, so we simply refer the reader to Ref.~\cite{selby2021accessible} for more discussion of this.
	
	In summary, accessible GPT fragments are simply GPT fragments, but represented in their native vector spaces, and closed under classical processing. 
	As with GPT fragments, the set of state vectors and effect covectors in an accessible GPT fragment need not be tomographically complete for each other (i.e., they need not span the same vector space and its dual), and the set of state vectors may contain subnormalized states whose normalised counterparts are not in the accessible GPT fragment.

	In order to formalize this move from the GPT fragments of the previous section to accessible GPT fragments, we will make use of the idempotent maps that we introduced. In particular, let us define a ``splitting'' of these idempotents as follows.
	In the case of $\Pi_{\Omega^\mathfrak{F}}$, this means finding a vector space $S_{\Omega^\mathfrak{A}}$ and a pair of linear maps
	\beq %
\InputIfFileExists{Diagrams/PProj.tikz}{}{\input{./figures/Diagrams/PProj.tikz}} :S \to S_{\Omega^\mathfrak{A}} \quad\text{and} \quad 
\InputIfFileExists{Diagrams/PInc.tikz}{}{\input{./figures/Diagrams/PInc.tikz}}:S_{\Omega^\mathfrak{A}}\to S
	\eeq
	such that
	\beq
\InputIfFileExists{Diagrams/StateProjection.tikz}{}{\input{./figures/Diagrams/StateProjection.tikz}}\ \  =\ \ %
\InputIfFileExists{Diagrams/PProjInc.tikz}{}{\input{./figures/Diagrams/PProjInc.tikz}}\quad\text{and}\quad  %
\InputIfFileExists{Diagrams/PIncProj.tikz}{}{\input{./figures/Diagrams/PIncProj.tikz}}\ \  =\ \  %
\InputIfFileExists{Diagrams/SPIdent.tikz}{}{\input{./figures/Diagrams/SPIdent.tikz}}.  
	\eeq
	In particular, these conditions mean that $S_{\Omega^\mathfrak{A}} \cong \mathsf{Span}[\Omega^\mathfrak{F}]$, so we will think of $S_{\Omega^\mathfrak{A}}$ as the vector space of states for the accessible GPT fragment. The map $P_{\Omega^\mathfrak{A}}$ is then a projector mapping states {\em viewed as vectors within the underlying GPT} to states {\em viewed as vectors in the accessible GPT fragment}. Meanwhile, the map $I_{\Omega^\mathfrak{A}}$ is an inclusion map taking states {\em viewed as vectors within the accessible GPT fragment} to states {\em viewed as vectors within the GPT}. Note that splitting an idempotent into a projector and inclusion map is unique up to some reversible transformation relating two different bases for $S_{\Omega^\mathfrak{A}}$. 
	
	The case of effects is handled in the same way, up to the caveat that we are thinking of all of the linear maps as acting contravariantly---that is, on the dual spaces, where the effects naturally live. That is, to split $\Pi_{\mathcal{E}^\mathfrak{F}}$, one finds a vector space $S_{\mathcal{E}^\mathfrak{A}}$ and a pair of linear maps
	\beq
\InputIfFileExists{Diagrams/MProj.tikz}{}{\input{./figures/Diagrams/MProj.tikz}}:S^* \to S_{\mathcal{E}^\mathfrak{A}}^*\quad \text{and} \quad %
\InputIfFileExists{Diagrams/MInc.tikz}{}{\input{./figures/Diagrams/MInc.tikz}}:S_{\mathcal{E}^\mathfrak{A}}^* \to S^*
	\eeq
	such that 
	\beq
\InputIfFileExists{Diagrams/EffectProjection.tikz}{}{\input{./figures/Diagrams/EffectProjection.tikz}} \ \ = \ \  %
\InputIfFileExists{Diagrams/MProjInc.tikz}{}{\input{./figures/Diagrams/MProjInc.tikz}}
	\quad\text{and}\quad %
\InputIfFileExists{Diagrams/MIncProj.tikz}{}{\input{./figures/Diagrams/MIncProj.tikz}}\ \  =\ \  %
\InputIfFileExists{Diagrams/SMIdent.tikz}{}{\input{./figures/Diagrams/SMIdent.tikz}}.
	\eeq
	Here we find that $S^*_{\mathcal{E}^\mathfrak{A}}\cong \mathsf{Span}[\mathcal{E}^\mathfrak{F}]$ and so we think of this as the dual vector space of effects for the accessible GPT fragment. The map $P_{\mathcal{E}^\mathfrak{A}}$ can then be though of as a projector mapping effects, viewed as covectors in the GPT, to effects, viewed as covectors in the accessible GPT fragment, and $I_{\mathcal{E}^\mathfrak{A}}$ as an inclusion mapping effects, viewed as covectors in the accessible GPT fragment, to effects, viewed as covectors in the GPT. The choice of which idempotents to split, and which projector and inclusion map to split them into, then amounts to nothing more than picking a particular basis with which to represent the states and effects (one basis for the underlying GPT, and another for the accessible GPT fragment). That is, all of these choices simply result in equivalent accessible GPT fragments, as discussed in Ref.~\cite{selby2021accessible}. 
	
	We can then define all of the components of the accessible GPT fragment in terms of the GPT fragment together with these projector and inclusion maps. In particular, we define the states and effects in the accessible GPT fragment as
	\beq
\InputIfFileExists{Diagrams/AGPTState.tikz}{}{\input{./figures/Diagrams/AGPTState.tikz}}\ \   :=\ \ %
\InputIfFileExists{Diagrams/accState.tikz}{}{\input{./figures/Diagrams/accState.tikz}} \quad \text{and} \quad
\InputIfFileExists{Diagrams/AGPTEffect.tikz}{}{\input{./figures/Diagrams/AGPTEffect.tikz}}\ \ := \ \      %
\InputIfFileExists{Diagrams/accEffect.tikz}{}{\input{./figures/Diagrams/accEffect.tikz}},
	\eeq
	and so, in particular, the unit effect for the accessible GPT fragment is given by
	\beq
\InputIfFileExists{Diagrams/discard.tikz}{}{\input{./figures/Diagrams/discard.tikz}}  \ \ :=\ \ %
\InputIfFileExists{Diagrams/accDiscard.tikz}{}{\input{./figures/Diagrams/accDiscard.tikz}}.
	\eeq
	Finally, we define the probability rule for the accessible GPT fragment by including the states and effects in the accessible GPT fragment into the underlying GPT and computing the probability within the underlying GPT via Eq.~\eqref{eq:GPTProbRule}. That is, we define a linear probability rule
	\beq
\InputIfFileExists{Diagrams/probRule.tikz}{}{\input{./figures/Diagrams/probRule.tikz}} \ \ :=\ \  %
\InputIfFileExists{Diagrams/accProbRule.tikz}{}{\input{./figures/Diagrams/accProbRule.tikz}},
	\eeq
	and use this to compute probabilities via
	\beq
	\mathrm{Prob}\left(%
\InputIfFileExists{Diagrams/AGPTEffect.tikz}{}{\input{./figures/Diagrams/AGPTEffect.tikz}},%
\InputIfFileExists{Diagrams/AGPTState.tikz}{}{\input{./figures/Diagrams/AGPTState.tikz}}\right)\quad :=\quad %
\InputIfFileExists{Diagrams/AGPTPRule.tikz}{}{\input{./figures/Diagrams/AGPTPRule.tikz}}.
	\eeq

	Succinctly, then, an accessible GPT fragment is specified by a quadruple
	\begin{align}\label{eq:def.accessible}
		&\hspace{-2ex}\mathfrak{A}=\left(\hspace{-0.5ex}\left\{%
\InputIfFileExists{Diagrams/AGPTState.tikz}{}{\input{./figures/Diagrams/AGPTState.tikz}} \hspace{-1ex}\right\}_{\hspace{-1ex} s\in \Omega^\mathfrak{A}}\hspace{-1ex},\  \left\{%
\InputIfFileExists{Diagrams/AGPTEffect.tikz}{}{\input{./figures/Diagrams/AGPTEffect.tikz}}\hspace{-1ex}\right\}_{\hspace{-1ex}  e\in \mathcal{E}^\mathfrak{A} }\hspace{-1ex},\  %
\InputIfFileExists{Diagrams/probRule.tikz}{}{\input{./figures/Diagrams/probRule.tikz}} ,\ %
\InputIfFileExists{Diagrams/discard.tikz}{}{\input{./figures/Diagrams/discard.tikz}} \hspace{-1ex}\right)\!\!.
	\end{align}
	In this paper, we will assume that the number of extreme points in  $\Omega^\mathfrak{A}$ and $\mathcal{E}^\mathfrak{A}$  are finite, as will be the case in any real experiment.
	
	In short, from the GPT fragment characterising a given experiment, one can construct the associated accessible GPT fragment as follows. First, one closes $\Omega^\mathfrak{F}$ and $\mathcal{E}^\mathfrak{F}$ under classical processings. Then, one reconceptualizes the states and effects as living in their native subspaces, namely in $S_{\Omega^\mathfrak{A}} \cong \mathsf{Span}[\Omega^\mathfrak{F}]$ and $S_{\mathcal{E}^\mathfrak{A}} \cong \mathsf{Span}[\mathcal{E}^\mathfrak{F}]$ rather than in the underlying GPT's vector space $S$.

	The notion of classical explainability for accessible GPT fragments is very closely related to that introduced above for GPTs and for GPT fragments. Since it is the main notion we use in this work, we discuss it in detail in the next section. It is straightforward to show (directly from the definitions and Eq.~\eqref{eq:FragmentPRuleEquiv}) that classical explainability of a GPT fragment is equivalent to classical explainability of the associated accessible GPT fragment.  This implies that one can work equally well with either the practical or the theoretical perspectives introduced above. Moreover, it implies that the particular choices that were made for the projection maps are irrelevant to the assessment of classicality (as one would expect).
	
	\subsection{Classical explainability of accessible GPT fragments}\label{sec:SEofAGPTF}
	
	One can now ask whether a given accessible GPT fragment is classically-explainable. The appropriate notion of classical-explainability for accessible GPT fragments, first introduced in Ref.~\cite{selby2021accessible}, is again a natural extension of the notion for standard GPTs:
	\begin{definition}[Simplicial-cone embeddings and simplex embeddings of an accessible GPT fragment]\label{def:SEofAGPTF}
		A \emph{simplicial-cone embedding}, $\tau_\mathfrak{A}$, of an accessible GPT fragment, $\mathfrak{A}$,
		is defined by a set of ontic states $\Lambda$ and a pair of linear maps
		\beq
\InputIfFileExists{Diagrams/iota.tikz}{}{\input{./figures/Diagrams/iota.tikz}} \quad\text{and}\quad %
\InputIfFileExists{Diagrams/kappa.tikz}{}{\input{./figures/Diagrams/kappa.tikz}}
		\eeq
		such that for all $s\in \Omega^\mathfrak{A}$ and for all $e\in \mathcal{E}^\mathfrak{A}$ we have
		\beq
\InputIfFileExists{Diagrams/IotaState.tikz}{}{\input{./figures/Diagrams/IotaState.tikz}} \ \geq_e\  0\quad\text{,}\quad %
\InputIfFileExists{Diagrams/KappaEffect.tikz}{}{\input{./figures/Diagrams/KappaEffect.tikz}} \ \geq_e\  0
		\eeq
		and such that
		\beq
\InputIfFileExists{Diagrams/probRule.tikz}{}{\input{./figures/Diagrams/probRule.tikz}}\quad=\quad %
\InputIfFileExists{Diagrams/probRuleSimplexDecomp.tikz}{}{\input{./figures/Diagrams/probRuleSimplexDecomp.tikz}}.
		\eeq
		
		A simplicial-cone embedding is said to be a \emph{simplex embedding} if it moreover satisfies
		\beq\label{eq:simplexconditiondiagramatic}
\InputIfFileExists{Diagrams/kappaCausal.tikz}{}{\input{./figures/Diagrams/kappaCausal.tikz}}\quad=\quad%
\InputIfFileExists{Diagrams/kappaCausal1.tikz}{}{\input{./figures/Diagrams/kappaCausal1.tikz}}.
		\eeq
	\end{definition}
	
	With these definitions in place it is straightforward to show an equivalence between classical-explainability of a GPT fragment and classical-explainability of the associated accessible GPT fragment. 
	\begin{prop}\label{prop:form1to2}
		Any simplex embedding for a GPT fragment $\mathfrak{F}$ (Def.~\ref{def:SEofGPTF}) can be converted into a simplex embedding for the associated accessible GPT fragment $\mathfrak{A}$ (Def.~\ref{def:SEofAGPTF}), and vice versa. 
	\end{prop}
	\proof
	Given a simplex embedding $(\fse,\fee)$ for the GPT fragment, one can construct a simplex embedding for the associated accessible GPT fragment by taking
	\beq
\InputIfFileExists{Diagrams/iota.tikz}{}{\input{./figures/Diagrams/iota.tikz}} :=%
\InputIfFileExists{Diagrams/newproof2.tikz}{}{\input{./figures/Diagrams/newproof2.tikz}}\quad \text{and} \quad %
\InputIfFileExists{Diagrams/kappa.tikz}{}{\input{./figures/Diagrams/kappa.tikz}} :=%
\InputIfFileExists{Diagrams/newproof3.tikz}{}{\input{./figures/Diagrams/newproof3.tikz}}
	\eeq
	The fact that this is a valid simplex embedding can be verified immediately from the definitions.
	
	Similarly, given a simplex embedding $(\ase,\aee)$ for the accessible GPT fragment, one can construct a simplex embedding for the GPT fragment by
	\beq\label{eq:APPexplcitAtoF}
\InputIfFileExists{Diagrams/iotaF.tikz}{}{\input{./figures/Diagrams/iotaF.tikz}} :=%
\InputIfFileExists{Diagrams/newproof1.tikz}{}{\input{./figures/Diagrams/newproof1.tikz}} \quad \text{and} \quad %
\InputIfFileExists{Diagrams/kappaF.tikz}{}{\input{./figures/Diagrams/kappaF.tikz}}:= %
\InputIfFileExists{Diagrams/newproof4.tikz}{}{\input{./figures/Diagrams/newproof4.tikz}}.
	\eeq
	Again, the fact that this is a valid simplex embedding can be verified immediately from the definitions.
	\endproof
	
	Next, we show that the existence of a simplicial-cone embedding (or a simplex embedding) can be checked via a linear program; see Section~\ref{app:Proof}.
	
	\section{Derivation of the linear program}\label{app:Proof}

	In this section we will show that, for any accessible GPT fragment, 
	simplicial cone embeddability can be tested with a linear program. Note that the linear program discussed in the main text is simply the special case where the accessible GPT fragment lives inside quantum theory, that is, when $S_{\Omega^\mathfrak{A}} \subseteq \mathsf{Herm}[\mathcal{H}]$ and $S_{\mathcal{E}^\mathfrak{A}} \subseteq \mathsf{Herm}[\mathcal{H}]$ for some $\mathcal{H}$. 
	
	Let $h_i$ be representative covectors for the extreme rays of the logical effect cone $\mathsf{Cone}[\Omega^\mathfrak{A}]^*$, and let $\n$ be the number of these extreme rays. By definition, each of these $h_i$ constitutes an inequality which defines a facet of the state cone  $\mathsf{Cone}[\Omega^\mathfrak{A}]$; conversely, every such facet is represented by some $h_i$. Then, one can define a map 
	\beq \label{eq:APPstatefacets}
\InputIfFileExists{Diagrams/HOmegaDef.tikz}{}{\input{./figures/Diagrams/HOmegaDef.tikz}}\quad:=\quad \sum_{i=1}^n \ \ %
\InputIfFileExists{Diagrams/HOmegaDef1.tikz}{}{\input{./figures/Diagrams/HOmegaDef1.tikz}}
	\eeq
	which takes any given state to the vector of $n$ values which it obtains on these $n$ facet inequalities.
	It follows that every vector $v$ is in the state cone if and only if it is mapped by $H_{\Omega^\mathfrak{A}}$ to a vector of positive values, i.e.,
	\beq\label{eq:stateChar}
\InputIfFileExists{Diagrams/facetCharStates.tikz}{}{\input{./figures/Diagrams/facetCharStates.tikz}} \ \ \geq_e\ 0 \quad \iff \quad %
\InputIfFileExists{Diagrams/facetCharStates1.tikz}{}{\input{./figures/Diagrams/facetCharStates1.tikz}} \ \ \in\ \mathsf{Cone}[ {\Omega^\mathfrak{A}}].
	\eeq
	Furthermore, any valid inequality satisfied by all vectors in the logical effect cone can be written as a positive linear sum of facet inequalities. Equivalently, one has that for any  $w \in \mathsf{Cone}[\Omega]^*$, there exists $\hat{w}\geq_e 0$ such that
	\beq \label{b3}
\InputIfFileExists{Diagrams/decompLogEff.tikz}{}{\input{./figures/Diagrams/decompLogEff.tikz}} \quad = \quad %
\InputIfFileExists{Diagrams/decompLogEff1.tikz}{}{\input{./figures/Diagrams/decompLogEff1.tikz}}.
	\eeq

	Similarly, let $g_j$ be representative vectors for the extreme rays of  $\mathsf{Cone}[{\mathcal{E}^\mathfrak{A}}]^*$. Let the number of such extreme rays be $m$, and define
	\beq \label{eq:APPeffectfacets}
\InputIfFileExists{Diagrams/HEDef.tikz}{}{\input{./figures/Diagrams/HEDef.tikz}}\quad:=\quad \sum_{j=1}^\m \ \ %
\InputIfFileExists{Diagrams/HEDef1.tikz}{}{\input{./figures/Diagrams/HEDef1.tikz}}.
	\eeq
	Then one has that
	\beq\label{eq:effectChar}
\InputIfFileExists{Diagrams/facetCharEffects.tikz}{}{\input{./figures/Diagrams/facetCharEffects.tikz}} \ \ \geq_e\ 0 \quad \iff \quad %
\InputIfFileExists{Diagrams/facetCharEffects1.tikz}{}{\input{./figures/Diagrams/facetCharEffects1.tikz}} \ \ \in\   \mathsf{Cone}[{\mathcal{E}^\mathfrak{A}}] 
	\eeq
	and that for any  $v\in \mathsf{Cone}[{\mathcal{E}^\mathfrak{A}}]^*$, there exists $\hat{v}\geq_e 0$ such that
	\beq
\InputIfFileExists{Diagrams/decompLogState.tikz}{}{\input{./figures/Diagrams/decompLogState.tikz}} \quad = \quad %
\InputIfFileExists{Diagrams/decompLogState1.tikz}{}{\input{./figures/Diagrams/decompLogState1.tikz}}.
	\eeq

	Recall that a simplicial-cone embedding is defined in terms of linear maps $\ase$ and $\aee$.  We now prove a useful lemma relating  $H_{\Omega^\mathfrak{A}}$ with $\ase$ and $H_{\mathcal{E}^\mathfrak{A}}$ with $\aee$. 
	
	\begin{lemma}\label{lem:iotakappaFactor}
		In any simplicial-cone embedding defined by linear maps $\ase$ and $\aee$, the map $\ase$ factors through  $H_{\Omega^\mathfrak{A}}$  as
		\beq
\InputIfFileExists{Diagrams/iota.tikz}{}{\input{./figures/Diagrams/iota.tikz}}\quad=\quad %
\InputIfFileExists{Diagrams/iotaDecomp.tikz}{}{\input{./figures/Diagrams/iotaDecomp.tikz}}
		\eeq
		and $\aee$ factors through  $H_{\mathcal{E}^\mathfrak{A}}$  as
		\beq
\InputIfFileExists{Diagrams/kappa.tikz}{}{\input{./figures/Diagrams/kappa.tikz}}\quad=\quad %
\InputIfFileExists{Diagrams/kappaDecomp.tikz}{}{\input{./figures/Diagrams/kappaDecomp.tikz}},
		\eeq
		where $\alpha: \mathds{R}^{\n} \to \mathds{R}^\Lambda$ and $\beta: \mathds{R}^\Lambda \to \mathds{R}^{\m} $ are matrices with nonnegative entries.
	\end{lemma}
	\proof
	Since  $\ase$ maps vectors in the state cone to points in the simplicial cone, it follows that for all $\lambda\in \Lambda$, one has
	\beq \label{b9}
\InputIfFileExists{Diagrams/lem2proof1.tikz}{}{\input{./figures/Diagrams/lem2proof1.tikz}} \quad \in \quad  \mathsf{Cone}[\Omega^\mathfrak{A}]^*. 
	\eeq
	To see that this is the indeed the case, note that Eq.~\eqref{b9}  asserts  that the process on its LHS is in the dual cone---i.e., it evaluates to a non-negative number on arbitrary vectors in the state cone. This is indeed the case, because if one composes an arbitrary vector in the state cone with $\ase$, one gets a vector in the simplicial cone by assumption; then, the effect $\lambda$ simply picks out the (necessarily non-negative) relevant coefficient corresponding to the $\lambda$ basis element.

	Hence, Eq.~\eqref{b3} implies that there exists a non-negative covector $v_\lambda$ such that
	\beq
\InputIfFileExists{Diagrams/lem2proof1.tikz}{}{\input{./figures/Diagrams/lem2proof1.tikz}} \quad=\quad %
\InputIfFileExists{Diagrams/lem2proof2.tikz}{}{\input{./figures/Diagrams/lem2proof2.tikz}} .
	\eeq
	Now, one simply inserts a resolution of the identity on the system coming out of $\ase$ and uses the above result to obtain the desired factorisation:
	\beq
\InputIfFileExists{Diagrams/lem2proof3.tikz}{}{\input{./figures/Diagrams/lem2proof3.tikz}} \quad =\quad %
\InputIfFileExists{Diagrams/lem2proof4.tikz}{}{\input{./figures/Diagrams/lem2proof4.tikz}} 
	\quad = \quad %
\InputIfFileExists{Diagrams/lem2proof5.tikz}{}{\input{./figures/Diagrams/lem2proof5.tikz}} 
	\quad =:\quad %
\InputIfFileExists{Diagrams/lem2proof6.tikz}{}{\input{./figures/Diagrams/lem2proof6.tikz}} .
	\eeq
	
	The proof for the factorisation of $\aee$ is almost identical, except that one inserts a resolution of the identity for the ingoing system rather than the outgoing system.
	\endproof
	
	With this in place, our main theorem is simple to prove. Linear Program 1 in the main text is a special case of this where the underlying GPT is taken to be quantum.  
	\begin{theorem}\label{thm:MainTheoremApp}
		Consider any accessible GPT fragment  ${\mathfrak{A}:= \{ \Omega^\mathfrak{A},\mathcal{E}^\mathfrak{A},B,\mathbf{u}\}}$ with state cone characterized by a matrix   $H_{\Omega^\mathfrak{A}}$  (whose codomain is dimension $n$)  and effect cone characterized by matrix $H_{\mathcal{E}^\mathfrak{A}}$  (whose domain is dimension $m$). Then the accessible GPT fragment $\mathfrak{A}$ is classically explainable if and only if  
		\beq \label{eq:b12}
		\exists\ \  %
\InputIfFileExists{Diagrams/sigma.tikz}{}{\input{./figures/Diagrams/sigma.tikz}}\ \ \geq_e\ 0\  \text{  such that}
		\eeq
		\beq \label{eq:b13}
\InputIfFileExists{Diagrams/probRule.tikz}{}{\input{./figures/Diagrams/probRule.tikz}} \quad=\quad  %
\InputIfFileExists{Diagrams/probRuleLPDecomp.tikz}{}{\input{./figures/Diagrams/probRuleLPDecomp.tikz}}.
		\eeq
	\end{theorem}

	\proof
	
	A simplicial-cone embedding is given by a $\aee$ and $\ase$  satisfying
	\beq\label{eq:SimplexDecomp}
\InputIfFileExists{Diagrams/probRule.tikz}{}{\input{./figures/Diagrams/probRule.tikz}}\quad=\quad %
\InputIfFileExists{Diagrams/probRuleSimplexDecomp.tikz}{}{\input{./figures/Diagrams/probRuleSimplexDecomp.tikz}}.
	\eeq
	If these exist, one can apply Lemma~\ref{lem:iotakappaFactor} to write
	\beq
\InputIfFileExists{Diagrams/probRule.tikz}{}{\input{./figures/Diagrams/probRule.tikz}} \quad=\quad %
\InputIfFileExists{Diagrams/probRuleSimplexDecomp.tikz}{}{\input{./figures/Diagrams/probRuleSimplexDecomp.tikz}} \quad = \quad %
\InputIfFileExists{Diagrams/maintheoremproof2.tikz}{}{\input{./figures/Diagrams/maintheoremproof2.tikz}} \quad =: \quad %
\InputIfFileExists{Diagrams/probRuleLPDecomp.tikz}{}{\input{./figures/Diagrams/probRuleLPDecomp.tikz}}.
	\eeq
	Hence, we arrive at a decomposition of the form of Eq.~\eqref{eq:b13}, where furthermore $\sigma \geq_e 0$, since $\alpha$ and $\beta$ are both entry-wise positive.
	
	Conversely, if there is a decomposition of the form given by Eq.~\eqref{eq:b13}, 
	then we can define $\aee$ and $\ase$  as
	\beq\label{eq:b16}
\InputIfFileExists{Diagrams/probRule.tikz}{}{\input{./figures/Diagrams/probRule.tikz}} \quad = \quad %
\InputIfFileExists{Diagrams/maintheoremproof1.tikz}{}{\input{./figures/Diagrams/maintheoremproof1.tikz}}  \quad=:\quad %
\InputIfFileExists{Diagrams/probRuleSimplexDecomp.tikz}{}{\input{./figures/Diagrams/probRuleSimplexDecomp.tikz}}
	\eeq
	to yield a valid simplicial-cone embedding. In particular, (i) $\aee$ and $\ase$ are clearly linear; (ii) $\ase$ maps states into the simplicial cone, since $H_{\Omega^\mathfrak{A}}$  satisfies Eq.~\eqref{eq:stateChar}; and (iii) $\aee$ maps effects into the dual of the simplicial cone, since $H_{\mathcal{E}^\mathfrak{A}}$  satisfies Eq.~\eqref{eq:effectChar} and $\sigma\geq_e 0$.
	\endproof
	
	The condition expressed in the statement of Theorem~\ref{thm:MainTheoremApp} provides us with our linear program for testing for classical-explainability of an accessible GPT fragment, or equivalently, classical-explainability of a GPT fragment from which it came. The core of the linear program is finding a suitable matrix $\sigma \geq_e 0$. 
	From a solution $\sigma$, one can construct a simplicial-cone embedding via Eq.~\eqref{eq:b16}. From this, one can construct a simplex embedding, which we do in the next section. In the section after that, we explicitly construct the ontological model for the GPT which is equivalent to the simplex-embedding.

	\subsection{From simplicial-cone embeddings to simplex embeddings}\label{app:NCOM}
	
	In Ref.~\cite{selby2021accessible}, we showed that if a simplicial-cone embedding exists then so too does a simplex embedding. Recall that the latter is just a simplicial-cone embedding satisfying an additional constraint on $\aee$, namely, that:
	\beq\label{eq:CausalProof}
\InputIfFileExists{Diagrams/kappaCausal.tikz}{}{\input{./figures/Diagrams/kappaCausal.tikz}}\quad=\quad%
\InputIfFileExists{Diagrams/kappaCausal1.tikz}{}{\input{./figures/Diagrams/kappaCausal1.tikz}}.
	\eeq
	We will now show how, given any simplicial-cone embedding given by $\ase'$ and $\aee'$ and ontic state space $\Lambda'$, we can construct a simplex embedding. This construction is useful because, by the results of Ref.~\cite{selby2021accessible}, it is equivalent to constructing an ontological model for the GPT fragment.
	
	The construction essentially removes superfluous ontic states and  then rescales the maps $\ase'$ and $\aee'$ (in a manner that can depend on the ontic state) to ensure that the representation of the ignoring operation is given by the all-ones vector, as Eq.~\eqref{eq:CausalProof} states. In what follows, we explain how this works.
	
	First, let us define $\Lambda:=\mathsf{Supp}[\tilde{u}]  \subseteq \Lambda'$---this will be the ontic state space for the simplex embedding---where $\tilde{u}$ is defined as
	\beq
\InputIfFileExists{Diagrams/kappaPrimeTrace1.tikz}{}{\input{./figures/Diagrams/kappaPrimeTrace1.tikz}}\quad:=\quad %
\InputIfFileExists{Diagrams/kappaPrimeTrace.tikz}{}{\input{./figures/Diagrams/kappaPrimeTrace.tikz}}\,.
	\eeq
	We can define a projection and an inclusion map for this subspace, which we denote as:
	\beq
\InputIfFileExists{Diagrams/inc.tikz}{}{\input{./figures/Diagrams/inc.tikz}}\quad\text{and}\quad %
\InputIfFileExists{Diagrams/proj.tikz}{}{\input{./figures/Diagrams/proj.tikz}}\,,
	\eeq
	respectively. Within this subspace, we can define an inverse of $\tilde{u}$, which we denote by $\tilde{u}^{-1}$, as the covector that satisfies:
	\beq\label{eq:pseudoinverse}
\InputIfFileExists{Diagrams/quasiInverse1.tikz}{}{\input{./figures/Diagrams/quasiInverse1.tikz}}\quad=\quad%
\InputIfFileExists{Diagrams/quasiInverse2.tikz}{}{\input{./figures/Diagrams/quasiInverse2.tikz}}.
	\eeq
	
	With these, we define the map $\aee$ (which describes the embedding of the effects) as
	\beq
\InputIfFileExists{Diagrams/kappa.tikz}{}{\input{./figures/Diagrams/kappa.tikz}}\quad := \quad %
\InputIfFileExists{Diagrams/simplexEmbeddingDef1.tikz}{}{\input{./figures/Diagrams/simplexEmbeddingDef1.tikz}}.
	\eeq
	This is just the required removal of ontic states (as dictated by the inclusion map) rescaling of $\aee'$ by the appropriate real values (as dictated by $\tilde{u}^{-1}$). These values are chosen to ensure that Eq.~\eqref{eq:CausalProof} is satisfied (as one can easily verify, as a direct consequence of Eq.~\eqref{eq:pseudoinverse}) and to ensure that it maps effects to entrywise-positive covectors. This is an explicit description of the rescaling matrix $R$ discussed in the main text.

	Then, we can define the map $\ase$ (that describes the simplex embedding of the states) as: 
	\beq
\InputIfFileExists{Diagrams/iota.tikz}{}{\input{./figures/Diagrams/iota.tikz}}\quad := \quad %
\InputIfFileExists{Diagrams/simplexEmbeddingDef.tikz}{}{\input{./figures/Diagrams/simplexEmbeddingDef.tikz}}.
	\eeq
	It is simple to verify that $\ase$ maps states to entrywise-positive vectors. All that then remains to be shown is that when $\ase$ and $\aee$ are composed that we reproduce the probability rule $B$.
	
	To see this, first note that
	\begin{align}
\InputIfFileExists{Diagrams/SEmbedP1.tikz}{}{\input{./figures/Diagrams/SEmbedP1.tikz}}\quad &= \quad %
\InputIfFileExists{Diagrams/SEmbedP2.tikz}{}{\input{./figures/Diagrams/SEmbedP2.tikz}}\\  = \quad  &%
\InputIfFileExists{Diagrams/SEmbedP3.tikz}{}{\input{./figures/Diagrams/SEmbedP3.tikz}}\quad = \quad %
\InputIfFileExists{Diagrams/SEmbedP4.tikz}{}{\input{./figures/Diagrams/SEmbedP4.tikz}}. \label{eq:c9}\end{align}
	Notice that the right-hand-side of Eq.~\eqref{eq:c9} 
	would be precisely equal to $B$ if it were not for the projection and inclusion maps in between $\ase$ and $\aee$. Hence, as a final step we need  to show that these maps are redundant in this expression. 
	
	To see this, first recall that for all $e\in\mathcal{E}^\mathfrak{A}$, there exists $e^\perp\in \mathcal{E}^\mathfrak{A}$ such that $e+e^\perp=\discardA$.
	Since $e$ and $e^\perp$ are both mapped to entrywise nonnegative covectors by $\aee'$, and because the sum of these covectors must be the vector $\tilde{u}$, it follows that 
	\beq
\InputIfFileExists{Diagrams/eSupp.tikz}{}{\input{./figures/Diagrams/eSupp.tikz}} \quad \in \ \mathsf{Supp}[\tilde{u}] = \Lambda \qquad  \forall \, e \in \mathcal{E}^\mathfrak{A}\,. 
	\eeq
	Hence, 
	\beq
\InputIfFileExists{Diagrams/eSupp.tikz}{}{\input{./figures/Diagrams/eSupp.tikz}}\quad=\quad %
\InputIfFileExists{Diagrams/eSupp2.tikz}{}{\input{./figures/Diagrams/eSupp2.tikz}}
	\eeq
	for all $e\in \mathcal{E}^\mathfrak{A}$. As $\mathcal{E}^\mathfrak{A}$ spans $S_{\mathcal{E}^\mathfrak{A}}$ we therefore have that:
	\beq
\InputIfFileExists{Diagrams/kappaPrime.tikz}{}{\input{./figures/Diagrams/kappaPrime.tikz}} \quad = \quad %
\InputIfFileExists{Diagrams/kappaPrimeProj.tikz}{}{\input{./figures/Diagrams/kappaPrimeProj.tikz}}.
	\eeq
	
	Putting this all together, we therefore have that
	\beq
\InputIfFileExists{Diagrams/probRuleSimplexDecomp.tikz}{}{\input{./figures/Diagrams/probRuleSimplexDecomp.tikz}}\quad = \quad %
\InputIfFileExists{Diagrams/SEDecomp.tikz}{}{\input{./figures/Diagrams/SEDecomp.tikz}} \quad=\quad  %
\InputIfFileExists{Diagrams/probRule.tikz}{}{\input{./figures/Diagrams/probRule.tikz}} 
	\eeq
	which completes the result.
	
	\subsection{From simplex embeddings to ontological models}\label{gettingncmodel}
	
	We have therefore seen how to transform a solution ($\sigma$) to the linear program into a simplicial-cone embedding, and from that to a simplex embedding. An explicit ontological model can be stated directly in terms of this embedding. Specifically, one defines the epistemic states and response functions in the ontological model as
	\beq
	\mu_s(\lambda) := \ \ %
\InputIfFileExists{Diagrams/epistemicstate.tikz}{}{\input{./figures/Diagrams/epistemicstate.tikz}}\quad \text{and} \quad \xi_e(\lambda) := \ \  %
\InputIfFileExists{Diagrams/responsefunction.tikz}{}{\input{./figures/Diagrams/responsefunction.tikz}}
	\eeq
	for all $\lambda \in \Lambda$, $s\in \Omega^\mathfrak{A}$, $e\in \mathcal{E}^\mathfrak{A}$. That this is a valid ontological model follows from the results of Refs.~\cite{schmid2021characterization,selby2021accessible}.

	\section{An operational measure of nonclassicality} \label{sec:opmeas}
	
	Thus far, we have only discussed the qualitative question of {\em whether or not} a classical explanation exists for a given scenario. A natural next question is to introduce quantitative measures of the {\em degree } of nonclassicality in one's scenario. A particularly useful approach to doing this would be to introduce a resource theory~\cite{coecke2016mathematical} of generalized noncontextuality and finding monotones~\cite{gonda2019monotones} therein. However, such an approach has not yet been developed. Therefore, here we take an approach motivated by the fact that {\em every} experiment admits of a classical explanation when subject to sufficient depolarizing noise~\cite{operationalks,marvian2020inaccessible}.  Hence, our approach is to quantify the robustness of one's nonclassicality---that is, the amount of noise which must be applied to one's data until it admits of a classical explanation. This is by no means a uniquely privileged measure, but it is operationally well-motivated. 
	
	There are many reasonable noise models, and which of these is most suitable depends on one's physical scenario. In this section, we show how one can adapt our linear program to quantify robustness of nonclassicality with respect to any noise model which treats noise as the probabilistic application of a 
	channel to all states in the experiment. That is, we consider arbitrary noise models of the form
	\beq\label{gennoisemod}
\InputIfFileExists{Diagrams/noise1.tikz}{}{\input{./figures/Diagrams/noise1.tikz}} := r\ %
\InputIfFileExists{Diagrams/depolarising1.tikz}{}{\input{./figures/Diagrams/depolarising1.tikz}} + (1-r)\ %
\InputIfFileExists{Diagrams/ident.tikz}{}{\input{./figures/Diagrams/ident.tikz}},
	\eeq
	where $N$ is an arbitrary channel (representing the noise).
	Note that Eq.~\eqref{gennoisemod} describes the noise as a channel in the full underlying GPT space $S$. One could alternatively describe noise as a channel acting on the spaces in which accessible GPT fragment lives (namely, a channel from $S_{\Omega^\mathfrak{A}}$ to $S_{\mathcal{E}^\mathfrak{A}}$); this is simply a further freedom in one's choice of noise model, and the techniques of this section apply to either approach.
	
	Each different noise model leads to a distinct  measure of robustness of nonclassicality. Perhaps the most common quantum noise models are those of this form and where $N$ is chosen to be either the completely depolarizing channel or the completely dephasing channel in a particular basis.
	For concreteness, in the quantum case, our open-source code is implemented assuming  depolarizing noise, as we detail below. For arbitrary GPTs, however, there is not necessarily a unique, well-defined maximally mixed state, and so the completely depolarising channel is not necessarily well-defined. In the GPT case, our open-source code therefore asks the user to specify a state to act as the maximally mixed state,  which  is then used to construct a completely depolarising channel, and the robustness to this noise channel is then computed and output by the program.  We explore the case of dephasing noise in Ref.~\cite{rossi2022contextuality}.

	Suppose the linear program discussed in Theorem~\ref{thm:MainTheoremApp} determines that a particular accessible GPT fragment does not admit of a simplex embedding. The natural next question  tackled in this section is then rephrased as  how much noise must one's experiment be subject to until it becomes simplex-embeddable. That is, what is the minimum value of $r$ for which one's experiment becomes classically-explainable?
	
	To address this question, we first translate Eq.~\eqref{gennoisemod} into its description at the level of the accessible GPT fragment, simply by applying the appropriate inclusion maps (and applying linearity of the inclusion map on the RHS):
	\beq
\InputIfFileExists{Diagrams/accNoisyProbRuleR.tikz}{}{\input{./figures/Diagrams/accNoisyProbRuleR.tikz}} := r\ %
\InputIfFileExists{Diagrams/accNoisyProbRule.tikz}{}{\input{./figures/Diagrams/accNoisyProbRule.tikz}} + (1-r)\ %
\InputIfFileExists{Diagrams/accProbRule.tikz}{}{\input{./figures/Diagrams/accProbRule.tikz}}.
	\eeq
	Then, we define
	\beq
\InputIfFileExists{Diagrams/probRuleN.tikz}{}{\input{./figures/Diagrams/probRuleN.tikz}}\ \ :=\ \ %
\InputIfFileExists{Diagrams/accNoisyProbRuleR.tikz}{}{\input{./figures/Diagrams/accNoisyProbRuleR.tikz}}\quad\text{and}\quad %
\InputIfFileExists{Diagrams/probRuleD.tikz}{}{\input{./figures/Diagrams/probRuleD.tikz}}\ \ :=\ \ %
\InputIfFileExists{Diagrams/accNoisyProbRule.tikz}{}{\input{./figures/Diagrams/accNoisyProbRule.tikz}},
	\eeq
	to write
	\beq
\InputIfFileExists{Diagrams/probRuleN.tikz}{}{\input{./figures/Diagrams/probRuleN.tikz}}  = r\ %
\InputIfFileExists{Diagrams/probRuleD.tikz}{}{\input{./figures/Diagrams/probRuleD.tikz}}  + (1-r)\ %
\InputIfFileExists{Diagrams/probRule.tikz}{}{\input{./figures/Diagrams/probRule.tikz}}.
	\eeq
	Hence, we see that the effect of the noise is simply to modify the linear map which captures the probability rule. 
	
	Clearly then, for a particular value of $r$, we can ask whether the new accessible GPT fragment, defined by replacing the old probability rule $B$ with this new one $B_{N_r}$, is simplex embeddable, simply by running the linear program. However, what is more interesting is to allow for $r$ to be an additional variable, and to ask: what is the minimal value of $r$ such that the associated accessible GPT fragment is simplex embeddable?
	
	This is formulated as the following optimization problem:
	\beq
	\mathsf{inf}\left\{  r\ \middle| \ \begin{array}{rl} r\ %
\InputIfFileExists{Diagrams/probRuleD.tikz}{}{\input{./figures/Diagrams/probRuleD.tikz}}  + (1-r)\ %
\InputIfFileExists{Diagrams/probRule.tikz}{}{\input{./figures/Diagrams/probRule.tikz}} &= \ \ %
\InputIfFileExists{Diagrams/probRuleLPDecomp.tikz}{}{\input{./figures/Diagrams/probRuleLPDecomp.tikz}}\\ \ %
\InputIfFileExists{Diagrams/sigma.tikz}{}{\input{./figures/Diagrams/sigma.tikz}} &\geq_e\ 0\\ r&\in\ [0,1]\end{array}\right\}, 
	\eeq
	which is also a linear program, since the only unknown quantities are the elements of $\sigma$ and $r$. 
	
	This linear program tells us the minimal amount of the noise channel $N$ which needs to be added to the experiment in order that it admit of a classical explanation. Similarly to before, the particular $\sigma$ that is found for the minimal value of $r$, can then be used to construct an explicit $\ase$ and $\aee$ which define the simplex embedding of the accessible GPT fragment that results after this amount of noise is applied.

	\section{Bounding the number of ontic states} \label{app:Bounds}
	
	Let us assume that an accessible GPT fragment $\mathcal{G}$ satisfies Theorem~\ref{thm:MainTheoremApp}, and hence we have a decomposition of the linear map $B$ as  
	\beq\label{eq:BinC}
\InputIfFileExists{Diagrams/probRule.tikz}{}{\input{./figures/Diagrams/probRule.tikz}}\quad=\quad %
\InputIfFileExists{Diagrams/probRuleLPDecomp.tikz}{}{\input{./figures/Diagrams/probRuleLPDecomp.tikz}} \quad = \quad \sum_{i,j=1}^{n,m} \sigma_{ij} %
\InputIfFileExists{Diagrams/cara1.tikz}{}{\input{./figures/Diagrams/cara1.tikz}},
	\eeq
	where $\sigma_{ij} \geq 0$. 
	
	It follows that $B$ belongs to a particular convex cone $\mathcal{C}\subset \mathcal{L}(S_{\Omega^\mathfrak{A}},S_{\mathcal{E}^\mathfrak{A}})$ living inside the real vector space of linear maps from $S_{\Omega^\mathfrak{A}}$ to $S_{\mathcal{E}^\mathfrak{A}}$; namely, the convex cone $\mathcal{C}$  given by the conic closure of a particular set of linear maps:
	\beq
	\left\{%
\InputIfFileExists{Diagrams/cara1.tikz}{}{\input{./figures/Diagrams/cara1.tikz}} \right\}_{i,j=1}^{n,m}.
	\eeq
	In other words,  $B$ belongs to the cone
	\beq
	\mathcal{C} := \left\{\ %
\InputIfFileExists{Diagrams/cara2.tikz}{}{\input{./figures/Diagrams/cara2.tikz}} \  \middle| \ %
\InputIfFileExists{Diagrams/cara2.tikz}{}{\input{./figures/Diagrams/cara2.tikz}} \ = \ \sum_{i,j=1}^{n,m}\gamma_{ij} \ %
\InputIfFileExists{Diagrams/cara1.tikz}{}{\input{./figures/Diagrams/cara1.tikz}} \ , \ \gamma_{ij}\geq 0 \right\},
	\eeq
	which is  clear from Eq.~\eqref{eq:BinC}. 
	
	We can therefore apply Carath\'eodory's theorem which, in this context, states that any linear map living inside the cone $\mathcal{C}$ can be decomposed as a conic combination of at most $\mathsf{dim}[\mathcal{L}(S_{\Omega^\mathfrak{A}},S_{\mathcal{E}^\mathfrak{A}})] = \mathsf{dim}[S_{\Omega^\mathfrak{A}}]\mathsf{dim}[S_{\mathcal{E}^\mathfrak{A}}]=:d_{{\Omega^\mathfrak{A}}} d_{\mathcal{E}^\mathfrak{A}}$ vertices of $\mathcal{C}$.
	
	Now, as we know that $B\in \mathcal{C}$, this means that there should exist coefficients $\chi_{ij}\geq 0$ such that the number of $\chi_{ij}\neq 0$ is at most $d_{{\Omega^\mathfrak{A}}} d_{\mathcal{E}^\mathfrak{A}}$. We write this new decomposition as
	\beq
\InputIfFileExists{Diagrams/probRule.tikz}{}{\input{./figures/Diagrams/probRule.tikz}} \quad = \quad \sum_{i,j=1}^{n,m} \chi_{ij} %
\InputIfFileExists{Diagrams/cara1.tikz}{}{\input{./figures/Diagrams/cara1.tikz}}.
	\eeq
	To make explicit that this decomposition only uses at most $d_{{\Omega^\mathfrak{A}}} d_{\mathcal{E}^\mathfrak{A}}$ non-zero elements, we switch to a single index $k\in\{1,...,d_{{\Omega^\mathfrak{A}}} d_{\mathcal{E}^\mathfrak{A}}\}=:K$ and write it as
	\beq
\InputIfFileExists{Diagrams/probRule.tikz}{}{\input{./figures/Diagrams/probRule.tikz}} \quad = \quad \sum_{k=1}^{d_{{\Omega^\mathfrak{A}}} d_{\mathcal{E}^\mathfrak{A}}} \tilde{\chi}_{k}  %
\InputIfFileExists{Diagrams/cara3.tikz}{}{\input{./figures/Diagrams/cara3.tikz}}\,, \eeq
	where $a$ and $b$ are functions such that $\tilde{\chi}_k=\chi_{a(k)b(k)}$. 
	
	Now it is simple to rewrite this into the form of a simplex embedding with ontic states indexed by $k\in K$:
	\begin{align}
\InputIfFileExists{Diagrams/probRule.tikz}{}{\input{./figures/Diagrams/probRule.tikz}} \quad &= \quad \sum_{k=1}^{d_{{\Omega^\mathfrak{A}}} d_{\mathcal{E}^\mathfrak{A}}} \tilde{\chi}_{k}  %
\InputIfFileExists{Diagrams/cara3.tikz}{}{\input{./figures/Diagrams/cara3.tikz}}\\
		&=\quad %
\InputIfFileExists{Diagrams/cara4n.tikz}{}{\input{./figures/Diagrams/cara4n.tikz}} \\
		&=:\quad   %
\InputIfFileExists{Diagrams/cara6n.tikz}{}{\input{./figures/Diagrams/cara6n.tikz}} \quad
		=:\quad %
\InputIfFileExists{Diagrams/probRuleSimplexDecompMin.tikz}{}{\input{./figures/Diagrams/probRuleSimplexDecompMin.tikz}}.
	\end{align}
	
	Thus, we have found a simplex embedding for the accessible GPT fragment into a simplex with vertices labeled by $K$. Consequently, the maximum number of ontic states that need be considered is $\mathsf{dim}[\mathcal{L}(S_{\Omega^\mathfrak{A}},S_{\mathcal{E}^\mathfrak{A}})] = \mathsf{dim}[S_{\Omega^\mathfrak{A}}]\mathsf{dim}[S_{\mathcal{E}^\mathfrak{A}}]=d_{{\Omega^\mathfrak{A}}} d_{\mathcal{E}^\mathfrak{A}}$. In the case of standard GPTs, where $S_{\Omega^\mathfrak{A}}=S_{\mathcal{E}^\mathfrak{A}}$ and $d_{{\Omega^\mathfrak{A}}}=d_{\mathcal{E}^\mathfrak{A}}=d$, the GPT dimension, we find that the maximal number of ontic states that need be considered is given by $d^2$. This is the same bound that was discovered in Ref.~\cite{gitton2020solvable}.

	It is unclear whether a tighter bound can be found. Certainly, in some specific examples, less than $d^2$ ontic states are required.
	Given a solution to the linear program, as a matrix $\sigma$, one can find a (potentially tighter) upper bound on the number of ontic states by finding the nonnegative rank of $\sigma$. The nonnegative rank is the minimal dimension of the vector space through which one can factor $\sigma$ in such a way that the two factors are nonnegative. It is easy to see that this defines an ontological model with a number of ontic states equal to the nonnegative rank, as the two factors of $\sigma$ can be used to define the $\ase$ and $\aee$ for a simplex embedding. There may, however, be many different $\sigma$ which are valid solutions to the linear program, and moreover, we do not have any proof that these necessarily have the same nonnegative rank. In order to find the minimal number of ontic states, one may therefore have to minimise the nonnegative rank over all possible solutions to the linear program. It is not clear whether there is an efficient method to solve this optimisation problem.

	\section{Worked Examples}\label{ap:workedex}
	
	In this section, we analyze the three illustrative examples we introduced in the main text  as well as one further example from the GPT known as Boxworld \cite{GPT_Barrett}.  We start with finite sets of states $\Omega$ and effects $\mathcal{E}$, and we show in detail how our linear program techniques can be used to assess their classicality. 
	In Section \ref{se:comparisonothers} of this Supplemental Material, we use these examples to comment on the related works in Refs.~\cite{gitton2020solvable,shahandeh2021contextuality}.
	
	\subsection{Example 1}
	
	Consider the set of four quantum states 
	\beq \label{stateex1}
	\Omega=\left\{ \ket{0}\bra{0} \,,\, \ket{1}\bra{1} \,,\, \ket{+}\bra{+} \,,\, \ket{-}\bra{-} \right\}\,,
	\eeq
	on a qubit.
	Consider moreover the finite set of effects 
	\beq \label{effectex1}
	\mathcal{E}=\left\{ \ket{0}\bra{0} \,,\, \ket{1}\bra{1} \,,\, \ket{+}\bra{+} \,,\, \ket{-}\bra{-} \,,\, \mathds{1}\,,\, 0 \right\}\,.
	\eeq
	Now the question is: {\em 
		are the statistics obtained by composing any state-effect pair classically-explainable?  }  
	To answer this question in the affirmative, we now show that these sets of states and effects satisfy with the requirements of Linear Program 1. 
	
	First, take the set of Hermitian operators  $O:=\left\{ \frac{1}{\sqrt2}\mathds{1}\,,\, \frac{1}{\sqrt2}\sigma_x\,,\, \frac{1}{\sqrt2}\sigma_z \right\}$,  where $\sigma_x$ and $\sigma_z$ are the Pauli-X and Pauli-Z operators, respectively.  
	Notice that $O$ is an orthonormal  basis for the subspace $S_\Omega$, and that in this basis the states in $\Omega$ are expressed as: \john
	\begin{align*}
		\ket{0}\bra{0} &\leftrightarrow\tfrac{1}{\sqrt2} \, [1,0,1]^T \,,\\
		\ket{1}\bra{1} & \leftrightarrow \tfrac{1}{\sqrt2} \, [1,0,-1]^T \,,\\
		\ket{+}\bra{+} &\leftrightarrow \tfrac{1}{\sqrt2} \, [1,1,0]^T \,,\\ 
		\ket{-}\bra{-} &\leftrightarrow \tfrac{1}{\sqrt2} \, [1,-1,0]^T \,.   
	\end{align*}\blk
	In this coordinate system, the Hermitian operators corresponding to the facet inequalities for  $\mathsf{Cone}[\Omega]$ (which has 4 facets) expressed in the basis $O$ read: 
	\begin{align*}
		h_1^\Omega &= [1,1,1]^T  \,, \\
		h_2^\Omega &=  [1,1,-1]^T  \,, \\
		h_3^\Omega &=  [1,-1,1]^T   \,, \\
		h_4^\Omega &=  [1,-1,-1]^T \,.
	\end{align*}
	Using these,
	we build the 
	linear map $H_\Omega$ that maps elements of $S_\Omega$ to vectors in $\mathds{R}^4$:
	\begin{equation}
		H_\Omega = \begin{bmatrix}
			1 & 1 & 1 \\
			1 & 1 & -1 \\
			1 & -1 & 1 \\
			1 & -1 & -1 
		\end{bmatrix} \,.
	\end{equation}
	For instance, 
	\beq
	H_\Omega[\ket{0}\bra{0}] =  \, [\sqrt2,0,\sqrt2,0]^T \,.
	\eeq
	
	Next, let us discuss the inclusion map for the case of states, $I_\Omega$. In this example we are working with a qubit system (that is, a quantum system in $\mathcal{H}_2$), hence $I_\Omega$ will map $S_\Omega$ into $\mathsf{Herm}[\mathcal{H}_2]$. Now, notice that our chosen basis for $S_\Omega$ is $O=\{ \tfrac{1}{\sqrt2}\mathds{1}\,,\, \tfrac{1}{\sqrt2}\sigma_x\,,\, \tfrac{1}{\sqrt2}\sigma_z \}$, while a basis for $\mathsf{Herm}[\mathcal{H}_2]$ is  $\{ \tfrac{1}{\sqrt2}\mathds{1}\,,\, \tfrac{1}{\sqrt2}\sigma_x\,,\, \tfrac{1}{\sqrt2}\sigma_y \,,\, \tfrac{1}{\sqrt2}\sigma_z \}$, where $\sigma_y$ is the Pauli-Y operator. Hence, the matrix representation of $I_\Omega$ in these bases is
	\beq
	I_\Omega = \begin{bmatrix}
		1 & 0 & 0 \\
		0 & 1 & 0 \\
		0 & 0 & 0 \\
		0 & 0 & 1 
	\end{bmatrix} \,.
	\eeq
	
	Similarly, one can define the corresponding bases, facet inequality operators, $H_\mathcal{E}$, and inclusion map for the set of effects $\mathcal{E}$. In a nutshell, notice that $O$ is also a basis for the subspace $S_\mathcal{E}$, hence:
\john	\begin{align*}
		0 &\leftrightarrow [0,0,0]^T\,,\\
		\ket{0}\bra{0} &\leftrightarrow\tfrac{1}{\sqrt2} \, [1,0,1]^T \,,\\
		\ket{1}\bra{1} &\leftrightarrow \tfrac{1}{\sqrt2} \, [1,0,-1]^T \,,\\
		\ket{+}\bra{+} &\leftrightarrow \tfrac{1}{\sqrt2} \, [1,1,0]^T \,,\\ 
		\ket{-}\bra{-} &\leftrightarrow \tfrac{1}{\sqrt2} \, [1,-1,0]^T \,,\\ \mathds{1} &\leftrightarrow [\sqrt2,0,0]^T  \,.
	\end{align*}\blk
	The facet-defining inequalities for $\mathsf{Cone}[\mathcal{E}]$ are hence the same as for $\mathsf{Cone}[\Omega]$. Using $\{h_1^\Omega,h_2^\Omega,h_3^\Omega,h_4^\Omega\}$ to define the linear map $H_\mathcal{E}$ that maps elements of $S_\mathcal{E}$ to vectors in $\mathds{R}^4$, one obtains
	\begin{equation}
		H_\mathcal{E} = H_\Omega \,.
	\end{equation}
	Finally, the inclusion map $I_\mathcal{E}$ embeds $S_\mathcal{E}$ into $\mathsf{Herm}[\mathcal{H}_2]$. Following the same argument as for the case of states, one hence obtains 
	\beq
	I_\mathcal{E} = I_\Omega\,.
	\eeq
	
	\
	
	Now that we have explicit forms for $H_\Omega$, $I_\Omega$, $H_\mathcal{E}$, and $I_\mathcal{E}$, we can use Linear Program 1 to assess the classical-explainability of the data generated by $(\Omega,\mathcal{E})$. More precisely, the statistics obtained by composing any state-effect pair is classically-explainable if there exists a matrix $\sigma$ with non-negative entries such that 
	\beq 
	I_\mathcal{E}^T\cdot I_\Omega
	= H_\mathcal{E}^T\cdot \sigma \cdot H_\Omega \,.
	\eeq
	One can then check that the following matrix does the job: 
	\beq
	\sigma = \tfrac{1}{4} \begin{bmatrix}
		1 & 0 & 0 & 0 \\
		0 & 1 & 0 & 0 \\
		0 & 0 & 1 & 0 \\
		0 & 0 & 0 & 1
	\end{bmatrix} = \tfrac{1}{4}\mathds{1}_4\,.
	\eeq
	
	\

	The simplex embedding is found from $\sigma$  and the $H$ matrices  (following Eq.~\eqref{eq:b16} and Section~\ref{app:NCOM})  to be given by
	\begin{align}
		\tau_\Omega = \tfrac{1}{2} \, \tau_\mathcal{E}\quad \text{and} \quad
		\tau_\mathcal{E}= \tfrac{1}{\sqrt2}
		\begin{bmatrix}
			1 & 1 & 1  \\
			1 & 1 & -1  \\
			1 & -1 & 1  \\
			1 & -1 & -1 
		\end{bmatrix} .
	\end{align}
	
	From this, one can find an ontological model for the scenario (following Section~\ref{gettingncmodel}). The model has four ontic states, over which the  epistemic states for the four quantum states in Eq.~\eqref{stateex1} are
	\begin{align}
		[\tfrac{1}{2}, 0, \tfrac{1}{2}, 0]^T, \quad
		[0,\tfrac{1}{2}, 0, \tfrac{1}{2}]^T, \quad
		[\tfrac{1}{2}, \tfrac{1}{2}, 0, 0]^T, \quad
		[0, 0, \tfrac{1}{2}, \tfrac{1}{2}]^T,
	\end{align}
	and the response functions for the six effects in Eq.~\eqref{effectex1} are
	\begin{align}\nonumber
		[1,0,1,0]^T,\quad
		&[0,1,0,1]^T,\quad
		[1,1,0,0]^T,\quad
		[0,0,1
		,1]^T,\\  
		&[1,1,1,1]^T, \quad
		[0,0,0,0]^T,
	\end{align}
	respectively. 
	One can directly check that this model reproduces the quantum statistics, and that it corresponds to the model in Fig. 1 of the main text.

	\subsection{Example 2}
	
	Here we consider the example of a quantum system of dimension four. The sets of states and effects that we consider
	are, respectively, 
	\beq \label{stateex2}
	\Omega=\{\ket{0}\bra{0}\,,\, \ket{1}\bra{1} \,,\, \ket{2}\bra{2}\,,\, \ket{3}\bra{3}\}\,,
	\eeq
	and
	\begin{align} \label{effectex2}
		\mathcal{E}=\{\ket{0}\bra{0}+\ket{1}\bra{1}\,,\,\ket{1}\bra{1}+\ket{2}\bra{2}\,,\,\ket{2}\bra{2}&+\ket{3}\bra{3}\,,\\ \nonumber
		&\ket{3}\bra{3}+\ket{0}\bra{0}\,,\, \mathds{1}_4\,,\, 0\}\,.    
	\end{align}
	Notice that although $\mathsf{Herm}[\mathcal{H}_4]$ is a 16-dimensional space,  $S_\Omega$ is 4-dimensional whilst $S_\mathcal{E}$ is only 3-dimensional. 
	
	Now, choose \john
	\begin{align*}
		&\{\tfrac{1}{2}\mathds{1}_4 \,,\,
		\tfrac{1}{2}(\ket{0}\bra{0}+\ket{1}\bra{1}- \ket{2}\bra{2}-\ket{3}\bra{3}) \,,\\
		&\tfrac{1}{2}(-\ket{0}\bra{0}+ \ket{1}\bra{1}+ \ket{2}\bra{2}- \ket{3}\bra{3}) \,,\\
		&\tfrac{1}{2}(\ket{0}\bra{0}-\ket{1}\bra{1}+ \ket{2}\bra{2}-\ket{3}\bra{3})\}\,,
	\end{align*} \blk
	as an othonormal basis of Hermitian operators for $S_\Omega$. Notice that in this representation, the four states that define $\Omega$ are associated with 4 vertices of a cube, and define a tetrahedron. Since the facets of $\mathsf{Cone}[\Omega]$ correspond to the facets of such tetrahedron, the number of facet-defining inequalities (i.e., rows of $H_\Omega$) is 4. 
	
	In addition, consider the following subbasis: \john
	\begin{align*} 
		&\{\tfrac{1}{2}\mathds{1}_4 \,,\,  
		\tfrac{1}{2}(\ket{0}\bra{0}+\ket{1}\bra{1}- \ket{2}\bra{2}-\ket{3}\bra{3}) \,,\\
		&\tfrac{1}{2}(-\ket{0}\bra{0}+\ket{1}\bra{1}+ \ket{2}\bra{2}-\ket{3}\bra{3})\}\,.
	\end{align*} \blk
	Linear combinations of these three Hermitian operators can actually yield the vertices of $\mathcal{E}$, and hence we can take them as a basis of Hermitian operators for $S_\mathcal{E}$ which is hence of dimension 3. Notice that in this representation, the four effects that define $\mathcal{E}$ correspond to the four vertices of a square, and hence the number of facets of $\mathsf{Cone}[\mathcal{E}]$ is 4. 
	
	Similarly to the previous example, one can then compute the linear maps $H_\Omega$ and $H_\mathcal{E}$, and here they are then found to be
	\beq
	H_\Omega = \begin{bmatrix}
		1 & 1 & 1 & -1 \\
		1 & 1 & -1 & 1 \\
		1 & -1 & 1 & 1 \\
		1 & -1 & -1 & -1 
	\end{bmatrix} \,,\quad 
	H_{\mathcal{E}}= \begin{bmatrix}
		1 & 1 & 1  \\
		1 & 1 & -1  \\
		1 & -1 & 1  \\
		1 & -1 & -1  \\
	\end{bmatrix} \,.
	\eeq
	On the other hand, notice that $\mathsf{Herm}[\mathcal{H}_4]$ is spanned by a Hermitian operator basis of 16 elements. The inclusion maps $I_\Omega$ and $I_\mathcal{E}$ then read
	\beq
	I_\Omega = \begin{bmatrix}
		\mathds{1}_4 \\
		\mathbf{0}_{12\times 4}
	\end{bmatrix}\,,\quad I_{\mathcal{E}} = \begin{bmatrix}
		\mathds{1}_3 \\
		\mathbf{0}_{13\times 3}
	\end{bmatrix}\,,
	\eeq
	where the matrix $\mathbf{0}_{a\times b}$ has dimension $a\times b$ and all entries equal to 0.
	
	\
	
	Now the question is whether the statistics from every pair of state-effect drawn from $(\Omega,\mathcal{E})$ can be explained classically. By taking the matrix 
	\beq
	\sigma=\tfrac{1}{4}\mathds{1}_4,
	\eeq
	one can straightforwardly check that $(\Omega,\mathcal{E})$ satisfy the condition of Linear Program 1, answering the question in the affirmative. 
	
	The simplex embedding is found from $\sigma$  and the $H$s  (following Eq.~\eqref{eq:b16} and Section~\ref{app:NCOM}) to be given by 
	\begin{align}
		\tau_\Omega = \tfrac{1}{2}
		\begin{bmatrix}
			1 & 1 & 1 & -1 \\
			1 & 1 & -1 & 1 \\
			1 & -1 & 1 & 1 \\
			1 & -1 & -1 & -1
		\end{bmatrix}, \quad \text{and} \quad
		\tau_\mathcal{E} = \tfrac{1}{2}\begin{bmatrix}
			1 & 1 & 1  \\
			1 & 1 & -1  \\
			1 & -1 & 1  \\
			1 & -1 & -1 
		\end{bmatrix} .
	\end{align}
	
	From this, one can find an ontological model for the scenario (following Section~\ref{gettingncmodel}). The model has four ontic states, over which the  epistemic states for the four quantum states in Eq.~\eqref{stateex2} are
	\begin{align}
		[0,1,0,0]^T, \quad
		[1,0,0,0]^T, \quad
		[0,0,1,0]^T, \quad
		[0,0,0,1]^T,
	\end{align}
	and the response functions for the six effects in Eq.~\eqref{effectex2} are
	\begin{align}\nonumber
		[1,1,0,0]^T,\quad [1,0,1,0]^T,\quad
		&\quad[0,0,1,1]^T,\quad
		[0,1,0,1]^T, \\
		&[1,1,1,1]^T, \quad
		[0,0,0,0]^T,
	\end{align}
	respectively.
	One can directly check that this model reproduces the quantum statistics, and that it corresponds to the model in Fig. 2 of the main text.
	
	\subsection{Example 3}\label{ap:newexample}
	
	Our third example has the same states and effects as our first example, except that the effects have been rotated by about the $\sigma_y$ axis by $\frac{\pi}{4}$. Explicitly, we consider the states
	\beq \label{stateex3new}
	\Omega=\left\{ \ket{0}\bra{0} \,,\, \ket{1}\bra{1} \,,\, \ket{+}\bra{+} \,,\, \ket{-}\bra{-} \right\}\,,
	\eeq
	and effects
	\beq \label{effectex3new}
	\mathcal{E}=\left\{ \mathcal{R}(\ket{0}\bra{0}) \,,\, \mathcal{R}(\ket{1}\bra{1}) \,,\, \mathcal{R}(\ket{+}\bra{+}) \,,\, \mathcal{R}(\ket{-}\bra{-}) \,,\, \mathds{1}\,,\, 0 \right\}\,
	\eeq
	where $\mathcal{R}(\_):=R_y(\frac{\pi}{4})(\_)R_y(-\frac{\pi}{4})$ and $R_y(\theta):=\left( \begin{smallmatrix}
		\cos(\frac{\theta}{2}) & \sin(\frac{\theta}{2}) \\
		-\sin(\frac{\theta}{2}) & \cos(\frac{\theta}{2}) 
	\end{smallmatrix}\right)$.
	This keeps us in the $\sigma_x-\sigma_z$ plane of the Bloch ball, and hence we can work with the same operator basis as in the first example.  As before, then,  we therefore obtain   
	\begin{equation}
		H_\Omega = \begin{bmatrix}
			1 & 1 & 1 \\
			1 & 1 & -1 \\
			1 & -1 & 1 \\
			1 & -1 & -1 
		\end{bmatrix} \,
	\end{equation}
	and
	\beq
	I_\Omega = I_\mathcal{E} = \begin{bmatrix}
		1 & 0 & 0 \\
		0 & 1 & 0 \\
		0 & 0 & 0 \\
		0 & 0 & 1 
	\end{bmatrix} \,.
	\eeq
	Unlike in the first example, however, the rotated effects lead to a different $H_\mathcal{E}$, namely
	\beq
	H_\mathcal{E} = \begin{bmatrix}
		13860 & 0 & 19601 \\
		13860 & 0 & -19601 \\
		13860 & 19601 & 0 \\
		13860 & -19601 & 0 
	\end{bmatrix} \,.
	\eeq
	Note that the code doesn't work with irrational numbers directly, so the above $H_\mathcal{E}$ is a rational approximation which works well for our purposes.
	
	Now that we have explicit forms for $H_\Omega$, $I_\Omega$, $H_\mathcal{E}$, and $I_\mathcal{E}$, we can use Linear Program 1 to assess the classical-explainability of the states and effects.
	
	We find that this set of states and effects is \emph{not} simplex embeddable, and hence is not classically explainable. Indeed, this remains the case until we depolarise by $r= \frac{8119}{27720}\sim 1 - \frac{1}{\sqrt{2}}$. Once we have depolarised by this amount, then we find that Linear Program 1 is satisfiable for
	\beq
	\sigma = \frac{1}{110880} \begin{bmatrix}
		1 & 0 & 1 & 0 \\
		0 & 1 & 0 & 1 \\
		1 & 1 & 0 & 0 \\
		0 & 0 & 1 & 1 
	\end{bmatrix} \, .
	\eeq
	
	The simplex embedding for the depolarised scenario can then be computed from $\sigma$  and the $H$ matrices  (following Eq.~\eqref{eq:b16} and Section~\ref{app:NCOM})  to be given by
	\begin{align}
		\ase = \frac{1}{2\sqrt{2}} 
		\begin{bmatrix}
			1 & 1 & 1 \\
			1 & 1 & -1 \\
			1 & -1 & 1 \\
			1 & -1 & -1 
		\end{bmatrix} \,\quad \text{and} \quad
		\aee = \frac{1}{\sqrt{2}}
		\begin{bmatrix}
			1 & \frac{1}{\sqrt{2}} & \frac{1}{\sqrt{2}} \\
			1 & \frac{1}{\sqrt{2}} & -\frac{1}{\sqrt{2}} \\
			1 & -\frac{1}{\sqrt{2}} & \frac{1}{\sqrt{2}}\\
			1 & -\frac{1}{\sqrt{2}} & -\frac{1}{\sqrt{2}} 
		\end{bmatrix} \, ,
	\end{align}
	From these, one can then find an ontological model for the depolarised scenario (following Section~\ref{gettingncmodel}). 
	
	Specifically, we find that the states in the depolarised scenario are represented as 
	\begin{align}
		[\tfrac{1}{2}, 0, \tfrac{1}{2}, 0]^T, \quad
		[0,\tfrac{1}{2}, 0, \tfrac{1}{2}]^T, \quad
		[\tfrac{1}{2}, \tfrac{1}{2}, 0, 0]^T, \quad
		[0, 0, \tfrac{1}{2}, \tfrac{1}{2}]^T,
	\end{align}
	and the response functions for the six effects are
	\begin{align}\nonumber
		[1,\tfrac{1}{2},\tfrac{1}{2},0]^T,\quad &[0,\tfrac{1}{2},\tfrac{1}{2},1]^T,\quad
		[\tfrac{1}{2},0,1,\tfrac{1}{2}]^T,\quad
		[\tfrac{1}{2},1,0,\tfrac{1}{2}]^T, \\
		&[1,1,1,1]^T, \quad
		[0,0,0,0]^T.
	\end{align}
	One can then directly check that this model reproduces the quantum statistics in the scenario depolarised by $r=1-\frac{1}{\sqrt{2}}$, and that it corresponds to the model in Fig. 3 of the main text.

	\subsection{Example 4 }\label{ap:boxw}
	
	This final example is quite different to the ones we have so far presented in the sense that it does not pertain to quantum states and effects. Instead, here we will apply our technique to assess the classicality of experiments performed with states and effects for systems in the generalised probabilistic theory known as Boxworld \cite{Barrett2005}.

	Consider a single Boxworld system. Now, take the finite set of states
	\beq \label{stateex3}
	\Omega=\{ [1,1,0]^T\,,\, [1,0,1]^T \,,\, [1,-1,0]^T \,,\, [1,0,-1]^T \}\,.
	\eeq
	Given the states we started with, it follows that $\mathsf{ConvHull}[\Omega]$ corresponds to the full state space of the  Boxworld system. 
	
	Similarly, consider the following finite set of effects
	\begin{align} \label{effectex3}
		\mathcal{E}=\{ \tfrac{1}{2} \, [1,-1,-1]^T\,,\, \tfrac{1}{2} \, [1,1,-1]^T \,,\, \tfrac{1}{2} \, [1,1,1]^T \,,\, &\tfrac{1}{2} \, [1,-1,1]^T \,, \\
		&[1,0,0]^T \,,\, [0,0,0]^T \, \}\,. \nonumber
	\end{align}
	Given the particular set of effects we started with, it follows that $\mathsf{Cone}[\mathcal{E}]$ is precisely the cone of effects corresponding to the Boxworld system. 
	
	Now we want to decide whether the statistics generated by every possible pair state-effect in $(\Omega,\mathcal{E})$ can admit a classical explanation; that is, we want to decide whether the statistics generated by the Boxworld system under any possible state preparation and effect can be explained by an underlying ontological model.  
	So let us apply our linear program technique from Linear Program 1. On the one hand, the linear maps $H_\Omega$ and $H_\mathcal{E}$ take the form:
	\begin{align}
		H_\Omega &= \begin{bmatrix}
			1 & 1 & 1  \\
			1 & 1 & -1  \\
			1 &-1 & 1  \\
			1 & -1 & -1  
		\end{bmatrix},
		\quad H_\mathcal{E} = \begin{bmatrix}
			1 & 0 & 1  \\
			1& 0 & -1  \\
			1 &1 & 0  \\
			1 &-1 & 0  
		\end{bmatrix}.
	\end{align}
	On the other hand, the inclusion maps $I_\Omega$ and $I_{\mathcal{E}}$ satisfy $I_\Omega = I_\mathcal{E} = \mathds{1}_3$. 
	
	Providing these as inputs in the linear program, one can find that there does not exist any $\sigma$ with non-negative entries such that 
	\beq
	\mathds{1}_3 = H_\mathcal{E}^T\cdot \sigma \cdot H_\Omega \,.
	\eeq
	
	The maximally mixed state for a Boxworld system is given by $\mu = [1,0,0]^T$. With this specification, our linear program computes that the minimal amount $r$ of completely depolarising noise which must be added to the experiment until it becomes simplex embeddable is $r=0.5$.
	That is, for the noisy scenario defined by the original set of effects, together with the original set of states but subjected to depolarizing noise\footnote{One could equally well take the original set of states and incorporate the noise into the effects instead.} with probability $r=0.5$, one finds that Linear Program 1 is satisfiable, for
	\begin{equation}
		\sigma = 
		\tfrac{1}{8} 
		\begin{bmatrix}
			1 & 0 & 1 & 0 \\
			0 & 1 & 0 & 1 \\
			1 & 1 & 0 & 0 \\
			0 & 0 & 1 & 1 
		\end{bmatrix}.
	\end{equation}

	The simplex embedding is found from $\sigma$ (following Eq.~\eqref{eq:b16} and Section~\ref{app:NCOM}) to be given by 
	\begin{align}
		\ase &= \tfrac{1}{4}
		\begin{bmatrix}
			1 & 1 & 1 \\
			1 & 1 & -1 \\
			1 & -1 & 1 \\
			1 & -1 & -1
		\end{bmatrix},\quad \text{and} \quad
		\aee &= \tfrac{1}{2}\begin{bmatrix}
			2 & 1 & 1 \\
			2 & 1 & -1 \\
			2 & -1 & 1 \\
			2 & -1 & -1
		\end{bmatrix}. 
	\end{align}
	
	From this, one can find an ontological model for the scenario (following Section~\ref{gettingncmodel}). The model has four ontic states, over which the  epistemic states for the four Boxworld states in Eq.~\eqref{stateex3} are
	\begin{align}
		[\tfrac{1}{2},\tfrac{1}{2},0,0]^T,\quad
		[\tfrac{1}{2},0,\tfrac{1}{2},0]^T,\quad
		[0,0,\tfrac{1}{2},\tfrac{1}{2}]^T,\quad
		[0,\tfrac{1}{2},0,\tfrac{1}{2}]^T,
	\end{align}
	and the response functions for the six effects in Eq.~\eqref{effectex3} are
	\begin{align}\nonumber
		[0,\tfrac{1}{2},\tfrac{1}{2},1]^T,\quad
		&[\tfrac{1}{2},1,0,\tfrac{1}{2}]^T,\quad
		[1,\tfrac{1}{2},\tfrac{1}{2},0]^T,\quad
		[\tfrac{1}{2},0,1,\tfrac{1}{2}]^T,\\
		&[1,1,1,1]^T,\quad
		[0,0,0,0]^T,  
	\end{align}
	respectively.
	One can directly check that this model reproduces the noisy Boxworld statistics (for $r=0.5$).

	\makebox[0pt][l]{%
		\hspace{-0.45cm}\begin{minipage}{0.48\textwidth}
			\begin{center}
				\noindent\[
				\begin{tikzpicture}[scale=1]
					\node at (-2,-4) {\footnotesize{(a) LACK of embedding of states}};
					\node[draw,fill, color=gray, shape=circle,scale=.2] (a) at (0,0) {};
					\node[draw,fill, color=black, shape=circle,scale=.2] (b) at (-4,0) {};
					\node[draw,fill, color=black, shape=circle,scale=.2] (c) at (1,3) {};
					\node[draw,fill, color=black, shape=circle,scale=.2] (d) at (1,-3) {};
					\draw (b) -- (c) -- (d) -- (b) ;
					\draw[dotted] (a) -- (b) ;
					\draw[dotted] (a) -- (c) ;
					\draw[dotted] (a) -- (d) ;
					\node (z) at ($(a.center)!0.5!(c.center)$) {};
					\node (p) at ($(c.center)!0.5!(d.center)$) {};
					\node (o) at ($(d.center)!0.5!(b.center)$) {};
					\node (m) at ($(a.center)!0.5!(b.center)$) {};
					\node (m1) at ($(m)!0.5!(z)$) {};
					\node (m2) at ($(z)!0.5!(p)$) {};
					\node (m3) at ($(p)!0.5!(o)$) {};
					\node (m4) at ($(o)!0.5!(m)$) {};
					\node [draw,fill, color=darkgreen, shape=circle,scale=.4] (v1) at ($(m1)!-0.5!(m3)$) {};
					\node [draw,fill, color=darkgreen, shape=circle,scale=.4] (v3) at ($(m3)!-0.5!(m1)$) {};
					\node [draw,fill, color=darkgreen, shape=circle,scale=.4] (v2) at ($(m2)!-0.5!(m4)$) {};
					\node [draw,fill, color=darkgreen, shape=circle,scale=.4] (v4) at ($(m4)!-0.5!(m2)$) {};
					\draw[thick, dashed, color=darkgreen] (v2) -- (v1) -- (v4) ;
					\draw[thick, color=darkgreen]  (v2) -- (v3) -- (v4) ;
					\path[color=darkgreen, fill, fill opacity = 0.5] (v1.center) -- (v2.center) -- (v3.center) -- (v4.center) -- (v1.center) -- cycle;
					\path [name path=v1--v4] (v4) -- (v1);
					\path [name path=d--b] (b) -- (d);
					\path [name intersections={of=v1--v4 and d--b,by=E}];
					\draw[thick, color=darkgreen]  (v4) -- (E) ;
					\path [name path=v1--v2] (v2) -- (v1);
					\path [name path=d--c] (c) -- (d);
					\path [name intersections={of=v1--v2 and d--c,by=F}];
					\draw[thick, color=darkgreen]  (v2) -- (F) ;
					\draw[dotted, color=gray!0.5!black] (o.center) -- (m.center) -- (z.center) -- (p.center) ;
					\draw[color=gray!0.5!black] (o.center) -- (p.center) ;
					\node [above left of = v1, node distance=0.35cm] () {$\textbf{S}_\textbf{1}$};
					\node [above of = v2, node distance=0.3cm] () {$\textbf{S}_\textbf{2}$};
					\node [below of = v3, node distance=0.25cm] () {$\textbf{S}_\textbf{3}$};
					\node [above left of = v4, node distance=0.35cm] () {$\textbf{S}_\textbf{4}$};
				\end{tikzpicture}
				\hspace{0.1cm}
				\begin{tikzpicture}[scale=0.8]
					\node[draw,fill, color=darkblue, shape=circle,scale=.4] (0101) at (-1.5,-0.5) {};
					\node[draw,fill, color=darkblue, shape=circle,scale=.4] (0011) at (0,0.5) {};
					\node[draw,fill, color=darkblue, shape=circle,scale=.4] (1100) at (2,-0.5) {};
					\node[draw,fill, color=darkblue, shape=circle,scale=.4] (1010) at (3.5,0.5) {};
					\node[draw,fill, color=darkblue, shape=circle,scale=.4] (1111) at (1,3.5) {};
					\node[draw,fill, color=darkblue, shape=circle,scale=.4] (0000) at (1,-3.5) {};
					\node [left of = 0101, node distance=0.35cm] () {\textbf{E}${}_\mathbf{4}$};
					\node [right of = 1010, node distance=0.35cm] () {\textbf{E}${}_\mathbf{2}$};
					\node [above right of = 0011, node distance=0.45cm] () {\textbf{E}${}_\mathbf{1}$};
					\node [below left of = 1100, node distance=0.42cm] () {\textbf{E}${}_\mathbf{3}$};
					\node [above of = 1111, node distance=0.3cm] () {$\mathds{1}$};
					\node [below of = 0000, node distance=0.3cm] () {$\mathbf{0}$};
					\draw[thick, color=darkblue] (1111) -- (0101) -- (0000) -- (1010) -- (1111) -- (1100) -- (0000) ;
					\draw[thick, color=darkblue] (0101) -- (1100) -- (1010) ;
					\draw[thick, dashed, color=darkblue] (1111) -- (0011) -- (0000) ;
					\draw[thick, dashed, color=darkblue] (0101) -- (0011) -- (1010) ;
					\path[color=darkblue, fill, fill opacity = 0.5] (1111.center) -- (1010.center) -- (0000.center) -- (0101.center) -- (1111.center) -- cycle;
					\node at (1,-5.5) {\footnotesize{(b) Embedding of effects}};
				\end{tikzpicture} 
				\]
			\end{center}
			
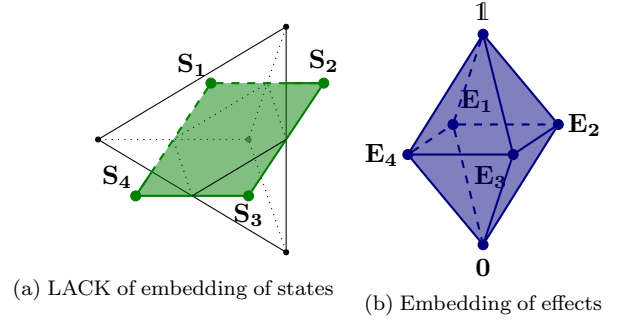
\captionof{figure}{ 
				\textbf{LACK of a classical explanation for Example 4.}
				(a) Depiction of the states in $\Omega$ (green dots), which fail to be embedded in a 3-dimensional slice of a 4-dimensional simplex. 
				(b) Depiction of the effects in $\mathcal{E}$ (blue dots), embedded in a 3-dimensional slice of the 4-dimensional hypercube that is dual to the simplex in (a).
				Note that the convex hull of the effects happens to cover the entire hypercube in this particular slice.
				Notice that for this particular choice of embedding of effects, that  the states must be represented (as shown in (a)) outside of the simplex, and so this does not constitute a simplex embedding. Because this is true for all possible embeddings of the effects, there is no possible noncontextual ontological model for---and hence classical explanation of---the operational scenario.
			}
			\label{fig:ome3}
		\end{minipage}
	}
	
	\section{Open-source code}
	
	This article is accompanied by a Mathematica™ notebook which automates the assessment of simplex embeddability. The supplied notebook has been tested only for Mathematica versions 12 and 13, and it requires an installation of the \textsf{cdd} binary, available for Ubuntu linux (and therefore for Windows via WSL) at \url{https://packages.ubuntu.com/jammy/amd64/libcdd-tools/filelist} and for Mac at \url{https://formulae.brew.sh/formula/cddlib}.

	\subsection{How to use the code }

	The code's main user-accessible function is \textsf{DiscoverEmbedding}. This function can take inputs in two different formats, depending on whether one's starting point is a fragment of quantum theory or a GPT fragment more generally. 
	That is, one may input 
	\begin{enumerate}[i.]
		\item  a set of density matrices and a set of POVM elements, or
		\item  a set of GPT states, a set of GPT effects, a specification of the GPT unit effect, and a specification of the maximally mixed state in the GPT.
	\end{enumerate} 
	
	If one is only interested in answering the {\em qualitative} question of whether or not a given GPT fragment is nonclassical, then the specification of the maximally mixed state in the GPT is irrelevant. (The code is written to reflect this---if one inputs a GPT fragment but does not specify any maximally mixed state, it will still output the value of $r$, which determines whether the scenario is classically-explainable or not.) 
	
	The formatting of these inputs is as follows. The function takes four arguments in the case of a GPT fragment, but only the first two are needed for the case of a quantum fragment. The four arguments are: a set of states, a set of effects, a unit effect, and a maximally mixed state. For the case of a quantum fragment, the set of states is an order-3 tensor, i.e., a comma-separated list of density operators as $n \times n$ complex-valued matrices. Similarly, the set of effects is then a list of POVM elements, each being an $n \times n$ complex-valued matrix. No unit effect need be specified for the case of a quantum fragment, as it is automatically presumed to be the $n \times n$ identity matrix. Similarly, the maximally mixed state is implicit when the input is a quantum fragment.
	
	For the case of a GPT fragment, one inputs a matrix $V_{\Omega^\mathfrak{F}}$ for the  set of states, a matrix $V_{\mathcal{E}^\mathfrak{F}}$ for the set of effects, a row  vector $u^\mathfrak{F}$ for the unit effect, and a row  vector $m^\mathfrak{F}$ for the maximally mixed state. The four arguments to the function \textsf{DiscoverEmbedding} are separated by commas.  $V_{\Omega^\mathfrak{F}}$
	should be constructed so that each of its rows is one of the GPT state vectors. Similarly, $V_{\mathcal{E}^\mathfrak{F}}$ should be constructed so that each of its rows is one of the GPT effect vectors.
	
	In both cases, the most important information returned by the code is the value of $r$. If $r=0$, then it follows that the scenario in question is classically-explainable, and the code returns such a classical explanation in the form of an ontological model. Namely, it returns a specification the epistemic states and the response functions which represent the states and effects (respectively) in one's input. By necessity~\cite{schmid2021characterization,schmid2020structure}, this ontological model constitutes a {\em noncontextual} ontological model for whatever operational scenario is described by the quantum or GPT input to the program.
	If $r>0$, then one has witnessed nonclassicality for one's operational scenario, and the value of $r$ is an operational measure of how nonclassical it is. The code again returns a noncontextual ontological model---not for the original scenario, but for the scenario after completely depolarizing noise is applied with probability $r$.

	The output ontological model is given as two matrices. The rows of the first matrix $\mu$ are the epistemic states, where the $i$th row is a representation of the $i$th quantum (GPT) state. The rows of the second matrix $\xi$ are the response functions, where the $i$th row is a representation of the $i$th quantum (GPT) effect.
	
	The code also outputs a number of other possibly useful objects. All of these are described in the following sections.

	\subsection{Useful matrix formulations}
	
	We now recast the diagrammatic results proven above in a language which is more amenable to coding, and so is useful for seeing how our proofs connect with our open-source code. Along the way, we highlight three equivalent formulations of simplicial-cone embedding and formulate these in terms of matrix computations. Respectively, these formulations are statements of
	\begin{enumerate}
		\item The existence of a simplicial-cone embedding for a GPT fragment.
		\item The existence of a simplicial-cone embedding for an accessible GPT fragment.
		\item The satisfaction of our linear program.
	\end{enumerate}
	
	We begin with a formulation that simply states the definition of simplicial-cone embeddability of a GPT fragment. Recall from the start of Section~\ref{sec:AccGPTFrag} that a GPT fragment is simply a subset $\Omega^\mathfrak{F}$ of states and a subset $\mathcal{E}^\mathfrak{F}$ of effects from some underlying GPT. Recall also that a simplicial-cone embedding for a fragment is the same as that for the underlying GPT, Def.~\ref{def:SimplexEmbeddingGPT}, but with the $\Omega^\mathcal{G}$ replaced by $\Omega^\mathfrak{F}$ and with $\mathcal{E}^\mathcal{G}$ replaced by $\mathcal{E}^\mathfrak{F}$.
	
	As a first step, as discussed above, we encode all of the states in $\Omega^\mathfrak{F}$ in a single matrix $V_{\Omega^\mathfrak{F}}^T$, which one defines by picking a basis for $S$, finding the coordinates of each GPT state vector in that basis, and then taking the resulting set of coordinate-vectors as the columns of $V_{\Omega^\mathfrak{F}}^T$.
	Analogously, define a single matrix $V_{\mathcal{E}^\mathfrak{F}}$ which characterizes the set of effects $\mathcal{E}^\mathfrak{F}$.
	
	Then, one can formulate simplicial-cone embedding of a GPT fragment as
	\begin{form}\label{rawform}{\textup{\bfseries [Compare to Definition~\ref{def:SEofGPTF}]}}
		\par\noindent Given a pair of matrices $V_{\Omega^\mathfrak{F}}$ and $V_{\mathcal{E}^\mathfrak{F}}$, 
		does there exist a pair of matrices $\fse$ and $\fee$ such that
		\begin{subequations}\begin{align}
				\fse \cdot V_{\Omega^\mathfrak{F}}^T &\geq_e  0\,, \\
				\fee\cdot V_{\mathcal{E}^\mathfrak{F}}^T &\geq_e 0\,, \\
				V_{\mathcal{E}^\mathfrak{F}}^T\cdot  V_{\Omega^\mathfrak{F}} &=V_{\mathcal{E}^\mathfrak{F}}^T\cdot \fee^T \cdot \fse \cdot  V_{\Omega^\mathfrak{F}} \,\text{?}
		\end{align}\end{subequations}
	\end{form}
	
	As shown in Proposition~\ref{prop:form1to2},
	a GPT fragment is simplicial-cone embeddable if and only if the associated {\em accessible} GPT fragment is simplicial-cone embeddable. 
	
	To write a matrix formulation of the simplicial-cone embeddability of an accessible GPT fragment, we introduce matrices $V_{\Omega^\mathfrak{A}}$ and $V_{\mathcal{E}^\mathfrak{A}}$ which collect together vector representations of states and effects in the accessible GPT fragment into a single matrix as we did for the case of GPT fragments. The operational probabilities are then computed according to the bilinear map ${B=I_{\mathcal{E}^\mathfrak{A}}^T \cdot I_{\Omega^\mathfrak{A}}}$ introduced in Section~\ref{sec:AccGPTFrag}. 
	
	We can then express simplicial-cone embeddability of accessible GPT fragments in terms of these matrices as
	\begin{form}\label{smalliotaform}{\textup{\bfseries [Compare to Definition~\ref{def:SEofAGPTF}]}}
		\par\noindent Given a pair of matrices $V_{\Omega^\mathfrak{A}}$ and $V_{\mathcal{E}^\mathfrak{A}}$ with a probability map $B$, does there exist a pair of matrices $\ase$ and $\aee$ such that 
		\begin{subequations}\begin{align}
				\ase \cdot
				V_{\Omega^\mathfrak{A}}^T &\geq_e 0\,, \\
				\aee\cdot 
				V_{\mathcal{E}^\mathfrak{A}}^T &\geq_e 0\,, \\
				B &= \aee^T \cdot \ase \,\text{?}
		\end{align}\end{subequations}
	\end{form}

	Our third and final formulation explicitly constitutes a linear program. Its equivalence to the previous formulation follows from Proposition~\ref{prop:form1to2}.

	In terms of matrices, it is given by
	\begin{form}{\textup{\bfseries [Compare to Theorem~\ref{thm:MainTheoremApp}]}}
		\par\noindent Given a pair of matrices $V_{\Omega^\mathfrak{A}}$ and $V_{\mathcal{E}^\mathfrak{A}}$ which compose to form probabilities according to the bilinear map ${B=I^T_{\mathcal{E}^\mathfrak{A}} \cdot I_{\Omega^\mathfrak{A}}}$,  
		does there exist any nonnegative matrix ${\sigma\geq_e 0}$ such that
		\begin{align}
			B = H_{\mathcal{E}^\mathfrak{A}}^T \cdot \sigma\cdot H_{\Omega^\mathfrak{A}}\,\text{?}
		\end{align}
	\end{form}
	
	All the formulations above refer to the existence of \emph{simplicial-cone} embeddings. We showed diagrammatically how to transform such an embedding into a \emph{simplex embedding} in  Appendix~\ref{app:NCOM}.  We now express this transformation in the language of this section. A simplex embedding is more restrictive than a simplicial-cone embedding in that it additionally requires that $\fee$ is such that $\fee\cdot {u}_\mathfrak{F}$ equals a vector (of length equal to the dimension of the simplex embedding) consisting of only ones. Given a \emph{simplicial-cone} embedding, one can readily construct a simplex embedding by rescaling row $i$ of the  simplicial-cone embedding's matrix $\fse$ by ${{\fee}[i]}^T\cdot{u}_\mathfrak{F}$, where $\fee[i]$ is the $i$-th column of $\fee$, and rescaling row $i$ of the simplicial-cone embedding's  $\fee$ by $\frac{1}{{{\fee}[i]}\cdot{u}_\mathfrak{F}}$, whenever ${{{\fee}[i]}\cdot{u}_\mathfrak{F}>0}$, and by truncating the vectors to remove the elements in which ${{{\fee}[i]}\cdot{u}_\mathfrak{F}=0}$.

	\subsection{Internals of the code}
	
	The function \textsf{DiscoverEmbedding} proceeds in four stages, at each stage printing useful information about the embeddability or nonembeddability of the quantum or GPT fragment. Moreover, there is an optional zeroth stage which is only relevant if the user provides a quantum fragment rather than a GPT fragment.
	
	\bigskip
	\noindent \textit{Stage 0 (when needed): Constructing a GPT fragment from a quantum fragment}\par
	If the input to the program is a specification of a set of quantum density matrices and a set of POVM elements, then there is a preliminary stage to the algorithm whereby the quantum fragment is recast in the GPT formalism. 
	
	First, we use the generalised Gell-Mann matrices~\cite{kimura2003bloch} as a basis for the space of Hermitian operators to represent the density matrices and POVM elements as real vectors. In particular, note that we choose the scaling of the Gell-Mann matrices so that the trace inner-product for Hermitian operators is mapped to the dot-product for the real vectors.
	Second, we compute the unit effect and maximally mixed state by converting the Hermitian operators $\mathds{1}$ and $\mathds{1}/d$ into  real vectors, again via the Gell-Mann matrices.
	
	This provides all of the data necessary for the second kind of input to the algorithm, and so the algorithm is then called on the input in this form. The rest of the stages are therefore the same regardless of which kind of input is given. 
	
	\bigskip
	\noindent \textit{Stage 1: Constructing an accessible GPT fragment from a GPT Fragment}\par
	The first task of our algorithm is to move from a GPT fragment to an \emph{accessible} GPT fragment; see Eq.~\ref{eq:def.accessible}. Thus, we initially construct (and print) the inclusion matrices $I_{\mathcal{E}^\mathfrak{A}}$ and $I_{\Omega^\mathfrak{A}}$, as well as the projection matrices $P_{\mathcal{E}^\mathfrak{A}}$ and $P_{\Omega^\mathfrak{A}}$ given by their respective Moore-Penrose pseudoinverses. The inclusion matrices are critical for constructing the left-hand side of condition (11b) of the main text in our linear program. The projection matrices are useful for converting  $\ase$ and $\aee$ to $\fse$ and $\fee$ respectively; see Eq.~\eqref{eq:APPexplcitAtoF}. Additionally, the projection matrices are used internally to construct an accessible GPT fragment, namely the three objects $V_{\Omega^\mathfrak{A}}$, $V_{\mathcal{E}^\mathfrak{A}}$, ${u^\mathfrak{A}}$, which are then printed for the user. 
	
	Internally, the inclusion and projection matrices are formed using Mathematica's \textsf{RowReduce} and \textsf{Pseudoinverse} commands. 
	
	\bigskip
	\noindent\textit{Stage 2: Computing the facets of the accessible GPT state and effect cones}\par
	Our linear program for characterizing the simplicial-cone embeddability of an accessible GPT fragment requires that we construct $H_{\Omega^\mathfrak{A}}$ and $H_{\mathcal{E}^\mathfrak{A}}$, e.g., for the right-hand side of Eq. (9b) of the main text. In particular, we define $H_{\Omega^\mathfrak{A}}$ and $H_{\mathcal{E}^\mathfrak{A}}$ as lists of the facet-defining inequalities, see Eqs. (2) and (6) of the main text or Eqs.~\eqref{eq:APPstatefacets} and \eqref{eq:APPeffectfacets}.
	
	The hypercone facet inequalities are derived internally by making an external call to the \textsf{cdd} binary.  Note that cdd relies on the double description method \cite{motzkin1953double} while in \cite{gitton2020solvable} they propose the use of the reverse search algorithm \cite{avis1991pivoting}. There are pros and cons to each of these algorithms, so in a future version we intend to allow the user to choose which to use. 
	
	The code then prints $H_{\Omega^\mathfrak{A}}$ and $H_{\mathcal{E}^\mathfrak{A}}$.  
	
	\bigskip
	\noindent\textit{Stage 3: Finding the robustness of nonclassicality}\par
	Next, the code implements the linear program described in Linear Program 2 in the main text.
	We explicitly return the minimum value of the primal objective, namely $r$, as well as
	a corresponding nonnegative matrix $\sigma$. To do so the maximally-mixed state is used to construct the depolarization map  $D^\mathfrak{F}\coloneqq  {m^\mathfrak{F}}^T\cdot {u^\mathfrak{F}}$. When no maximally-mixed state is explicitly provided by the user---or, in the case where the input is quantum, computed, as in stage 0---the code will automatically generate a ``central" state by taking the uniform mixture of all states in $V_{\Omega^\mathfrak{F}}$.  
	Recall that the particular noise model implemented by our code depends on the choice of $m^\mathfrak{F}$, and so the precise value of $r$ is only physically meaningful relative to this noise model; see again the discussion in Section~\ref{sec:opmeas}. For any choice, however, the input GPT fragment is classical simplex-cone embeddable if and only if $r=0$.
	
	If our program finds that the smallest $r$ is strictly positive, the code will further print out the matrix formed by the left-hand side (equivalently, by the right-hand side) of Eq. (11b) of the main text. (This information may be of interest insofar as it encodes the effective probability rule for the accessible GPT fragment generated after the depolarization.)

	\bigskip
	\noindent\textit{Stage 4: Conversion to a simplex embedding from a simplicial-cone embedding}\par
	Finally, the algorithm returns matrices for both $\ase$ and $\aee$, rescaled so that $\aee\cdot {u^\mathfrak{A}} = \left[1,1,...,1\right]$. That is, the final printout from the code is an explicit simplex embedding of the (depolarized, if necessary) accessible GPT fragment, per Eq.~\eqref{eq:simplexconditiondiagramatic}.
	
	Note that if $r>0$, the final simplex embedding given by $\ase$ and $\aee$ is actually not a simplex embedding of the \emph{original} accessible GPT fragment, but rather it is a simplex embedding of the \emph{minimally depolarized} accessible GPT fragment which \emph{is} simplicial-cone embeddable. 
	Recall that per Proposition~\ref{prop:form1to2} and Appendix~\ref{app:NCOM}, an accessible GPT fragment is simplex embeddable whenever it is simplicial-cone embeddable, so there is not loss of generality in returning simplex embeddings as opposed to simplicial-cone embeddings.

	Finally, an explicit ontological model for one's scenario is constructed, following the construction in Section~\ref{gettingncmodel}. This model is specified as a matrix $\mu$ whose rows form the epistemic states representing the states in (the rows of) $V_{\Omega^\mathfrak{F}}$, together with a matrix $\xi$ whose rows form the response functions representing the effects in (the rows of) $V_{\mathcal{E}^\mathfrak{F}}$.

\renewcommand{\addcontentsline}[2][]{\nocontentsline#1{#2}}

\end{document}

%% file: Diagrams/discardGPT.tikz
\begin{tikzpicture}
	\begin{pgfonlayer}{nodelayer}
		\node [style=none] (0) at (0, -0.5) {};
		\node [style=none] (1) at (0, 0.25) {};
		\node [style=right label] (3) at (0, -0.25) {$S$};
		\node [style=upground] (4) at (0, 0.5) {};
	\end{pgfonlayer}
	\begin{pgfonlayer}{edgelayer}
		\draw [qWire] (1.center) to (0.center);
	\end{pgfonlayer}
\end{tikzpicture}

%% file: Diagrams/discard.tikz
\begin{tikzpicture}
	\begin{pgfonlayer}{nodelayer}
		\node [style=none] (0) at (0, -0.5) {};
		\node [style=none] (1) at (0, 0.25) {};
		\node [style=right label] (3) at (0, -0.25) {$S_{\mathcal{E}^\mathfrak{A}}$};
		\node [style=upground] (4) at (0, 0.5) {};
	\end{pgfonlayer}
	\begin{pgfonlayer}{edgelayer}
		\draw [qWire] (1.center) to (0.center);
	\end{pgfonlayer}
\end{tikzpicture}

%% file: Diagrams/GPTState.tikz
\begin{tikzpicture}
	\begin{pgfonlayer}{nodelayer}
		\node [style=none] (0) at (0, 1) {};
		\node [style=point] (1) at (0, -0.25) {$s$};
		\node [style=right label] (3) at (0, 0.75) {$S$};
	\end{pgfonlayer}
	\begin{pgfonlayer}{edgelayer}
		\draw [qWire] (1) to (0.center);
	\end{pgfonlayer}
\end{tikzpicture}

%% file: Diagrams/GPTEffect.tikz
\begin{tikzpicture}
	\begin{pgfonlayer}{nodelayer}
		\node [style=none] (0) at (0, -1) {};
		\node [style=copoint] (1) at (0, 0.25) {$e$};
		\node [style=right label] (3) at (0, -0.75) {$S$};
	\end{pgfonlayer}
	\begin{pgfonlayer}{edgelayer}
		\draw [qWire] (1) to (0.center);
	\end{pgfonlayer}
\end{tikzpicture}

%% file: Diagrams/kappaCausal1.tikz
\begin{tikzpicture}
	\begin{pgfonlayer}{nodelayer}
		\node [style=none] (0) at (0, 0) {};
		\node [style=right label] (2) at (0, -0.75) {$\mathds{R}^\Lambda$};
		\node [style=none] (6) at (0, -0.75) {};
		\node [style=upground] (7) at (0, 0.25) {};
	\end{pgfonlayer}
	\begin{pgfonlayer}{edgelayer}
		\draw [cWire] (0.center) to (6.center);
	\end{pgfonlayer}
\end{tikzpicture}

%% file: Diagrams/State.tikz
\begin{tikzpicture}
	\begin{pgfonlayer}{nodelayer}
		\node [style=none] (0) at (0, 0.75) {};
		\node [style=point] (1) at (0, -0.75) {$s$};
	\end{pgfonlayer}
	\begin{pgfonlayer}{edgelayer}
		\draw [qWire] (0.center) to (1);
	\end{pgfonlayer}
\end{tikzpicture}

%% file: Diagrams/Effect.tikz
\begin{tikzpicture}
	\begin{pgfonlayer}{nodelayer}
		\node [style=none] (0) at (0, -0.75) {};
		\node [style=copoint] (1) at (0, 0.75) {$e$};
	\end{pgfonlayer}
	\begin{pgfonlayer}{edgelayer}
		\draw [qWire] (0.center) to (1);
	\end{pgfonlayer}
\end{tikzpicture}

%% file: Diagrams/SPIdent.tikz
\begin{tikzpicture}
	\begin{pgfonlayer}{nodelayer}
		\node [style=none] (2) at (0, 1) {};
		\node [style=none] (3) at (0, -1) {};
		\node [style=right label] (4) at (0, -0.25) {$S_{\Omega^\mathfrak{A}}$};
	\end{pgfonlayer}
	\begin{pgfonlayer}{edgelayer}
		\draw [qWire] (2.center) to (3.center);
	\end{pgfonlayer}
\end{tikzpicture}

%% file: Diagrams/SMIdent.tikz
\begin{tikzpicture}
	\begin{pgfonlayer}{nodelayer}
		\node [style=none] (2) at (0, 1) {};
		\node [style=none] (3) at (0, -1) {};
		\node [style=right label] (4) at (0, -0.25) {$S_{\mathcal{E}^\mathfrak{A}}$};
	\end{pgfonlayer}
	\begin{pgfonlayer}{edgelayer}
		\draw [qWire] (2.center) to (3.center);
	\end{pgfonlayer}
\end{tikzpicture}

%% file: Diagrams/AGPTState.tikz
\begin{tikzpicture}
	\begin{pgfonlayer}{nodelayer}
		\node [style=none] (0) at (0, 1) {};
		\node [style=point] (1) at (0, -0.25) {$s$};
		\node [style=right label] (3) at (0, 0.75) {$S_{\Omega^\mathfrak{A}}$};
	\end{pgfonlayer}
	\begin{pgfonlayer}{edgelayer}
		\draw [qWire] (1) to (0.center);
	\end{pgfonlayer}
\end{tikzpicture}

%% file: Diagrams/AGPTEffect.tikz
\begin{tikzpicture}
	\begin{pgfonlayer}{nodelayer}
		\node [style=none] (0) at (0, -1) {};
		\node [style=copoint] (1) at (0, 0.25) {$e$};
		\node [style=right label] (3) at (0, -0.75) {$S_{\mathcal{E}^\mathfrak{A}}$};
	\end{pgfonlayer}
	\begin{pgfonlayer}{edgelayer}
		\draw [qWire] (1) to (0.center);
	\end{pgfonlayer}
\end{tikzpicture}

%% file: Diagrams/facetCharStates1.tikz
\begin{tikzpicture}
	\begin{pgfonlayer}{nodelayer}
		\node [style=none] (13) at (0, 0.75) {};
		\node [style=point] (14) at (0, -0.5) {$v$};
		\node [style=right label] (17) at (0, 0.25) {$S_{\Omega^\mathfrak{A}}$};
	\end{pgfonlayer}
	\begin{pgfonlayer}{edgelayer}
		\draw [qWire] (14) to (13.center);
	\end{pgfonlayer}
\end{tikzpicture}

%% file: Diagrams/decompLogEff.tikz
\begin{tikzpicture}
	\begin{pgfonlayer}{nodelayer}
		\node [style=none] (0) at (0.5, -0.5) {};
		\node [style=copoint, fill=black] (1) at (0.5, 0.75) {\color{white}$w$};
		\node [style=right label] (2) at (0.5, 0) {$S_{\Omega^\mathfrak{A}}$};
	\end{pgfonlayer}
	\begin{pgfonlayer}{edgelayer}
		\draw [qWire] (1) to (0.center);
	\end{pgfonlayer}
\end{tikzpicture}

%% file: Diagrams/facetCharEffects1.tikz
\begin{tikzpicture}
	\begin{pgfonlayer}{nodelayer}
		\node [style=none] (13) at (0, -0.5) {};
		\node [style=copoint] (14) at (0, 0.75) {$w$};
		\node [style=right label] (17) at (0, 0) {$S_{\mathcal{E}^\mathfrak{A}}$};
	\end{pgfonlayer}
	\begin{pgfonlayer}{edgelayer}
		\draw [qWire] (14) to (13.center);
	\end{pgfonlayer}
\end{tikzpicture}

%% file: Diagrams/decompLogState.tikz
\begin{tikzpicture}
	\begin{pgfonlayer}{nodelayer}
		\node [style=none] (0) at (0.5, 0.75) {};
		\node [style=point, fill=black] (1) at (0.5, -0.5) {\color{white}$v$};
		\node [style=right label] (2) at (0.5, 0.5) {$S_{\mathcal{E}^\mathfrak{A}}$};
	\end{pgfonlayer}
	\begin{pgfonlayer}{edgelayer}
		\draw [qWire] (1) to (0.center);
	\end{pgfonlayer}
\end{tikzpicture}

%% file: Diagrams/kappaPrimeTrace1.tikz
\begin{tikzpicture}
	\begin{pgfonlayer}{nodelayer}
		\node [style=copoint] (0) at (0, 0) {$\tilde{u}$};
		\node [style=right label] (2) at (0, -0.75) {$\mathds{R}^{\Lambda'}$};
		\node [style=none] (6) at (0, -1) {};
	\end{pgfonlayer}
	\begin{pgfonlayer}{edgelayer}
		\draw [cWire] (0) to (6.center);
	\end{pgfonlayer}
\end{tikzpicture}

%% file: Diagrams/quasiInverse2.tikz
\begin{tikzpicture}
	\begin{pgfonlayer}{nodelayer}
		\node [style=none] (7) at (0, -0.25) {};
		\node [style=none] (11) at (0, -1.25) {};
		\node [style=right label] (12) at (0, -1) {$\mathds{R}^{\Lambda}$};
		\node [style=upground] (13) at (0, 0) {};
	\end{pgfonlayer}
	\begin{pgfonlayer}{edgelayer}
		\draw[cWire] (7.center) to (11.center);
	\end{pgfonlayer}
\end{tikzpicture}